\documentclass[12pt]{article}

\usepackage{authblk}
\usepackage{graphicx}
\usepackage{fullpage}
\usepackage{amsfonts}
\usepackage{booktabs}
\usepackage[version=4]{mhchem}

\title{\Large \bf Properties of cyanobacterial UV-absorbing pigments suggest their evolution was driven by optimizing photon dissipation rather than photoprotection}

\author[1]{Aleksandar Simeonov}
\author[2]{Karo Michaelian}

\affil[1]{Institute of Biology, Faculty of Natural Sciences and Mathematics, "University Ss. Cyril and Methodius", 1000 Skopje, Republic of North Macedonia}
\affil[2]{Department of Nuclear Physics and Application of Radiation, Institute of Physics, UNAM, Circuito Interior de la Investigaci\'{o}n Cient\'{i}fica, Cuidad Universitaria, M\'{e}xico D.F., Mexico, C.P. 04510}

\date{}

\begin{document}
 \maketitle
 
\begin{abstract}
An ancient repertoire of ultraviolet (UV)-absorbing pigments which survive today in the phylogenetically oldest extant photosynthetic organisms, the cyanobacteria, point to a direction in evolutionary adaptation of the pigments and their associated biota; from largely UVC-absorbing pigments in the Archean to pigments covering ever more of the longer wavelength UV and visible in the Phanerozoic. Such a scenario implies selection of photon dissipation rather than photoprotection over the evolutionary history of life. This is consistent with the thermodynamic dissipation theory of the origin and evolution of life which suggests that the most important hallmark of biological evolution has been the covering of Earth's surface with organic pigment molecules and water to absorb and dissipate ever more completely the prevailing surface solar spectrum. In this article we compare a set of photophysical, photochemical, biosynthetic and other germane properties of the two dominant classes of cyanobacterial UV-absorbing pigments, the mycosporine-like amino acids (MAAs) and scytonemins. Pigment wavelengths of maximum absorption correspond with the time dependence of the prevailing Earth surface solar spectrum, and we proffer this as evidence for the selection of photon dissipation rather than photoprotection over the history of life on Earth.
\end{abstract}

\clearpage

\section{Introduction}
Once the subject of mystical and metaphysical interpretations, the explanation of life on Earth has slowly gained a physical-chemical grounding in biochemistry and non-equilibrium thermodynamics. Founded on Boltzmann's nineteenth century insights into thermodynamics, then further elaborated by twentieth century scientists, notably by Ilya Prigogine, non-equilibrium thermodynamics attempts to explain the phenomenon of life as ``dissipative structuring''; an out of equilibrium organization of matter in space and time under an impressed external potential for the explicit purpose of producing entropy (Prigogine, 1967; Glansdorff and Prigogine, 1971). 

Using the formalism of Prigogine's ``Classical Irreversible Thermodynamics'' in the non-linear regime, Michaelian (2009; 2011; 2012; 2013; 2016; 2017) has proposed a theory for life's origin and evolution as microscopic self-organized dissipative structuring of organic pigment molecules and the dispersal of these over the entire surface of the Earth as a response to the impressed high-energy (UVC to visible) solar photon spectrum prevailing at Earth's surface. All physicochemical structuring associated with the pigments, such as the photosynthetic organisms primarily, and heterotrophic organisms secondarily, can be regarded as agents for the synthesis, proliferation and distribution of the pigments. The theory suggests that it is the thermodynamic imperative of increasing the entropy production of Earth in its solar environment that is behind the vitality of living matter as seen in its ability to proliferate, diversify, and evolve. 

The theory explains satisfactorily, for example, why the three major classes of photosynthetic pigments (chlorophylls, carotenoids and phycobilins) of phototrophic organisms dissipate most of the absorbed photonic energy into heat (a process known as non-photochemical quenching, NPQ) while funneling only a minute fraction into productive photochemistry (Horton et al., 1996; Ruban et al., 2007; Staleva et al., 2015; Gupta et al., 2015). Moreover, these organisms often contain a vast array of other organic pigments in addition to the photosynthetic ones, whose absorption spectra extend outside of the photosynthetically active radiation (PAR) in the visible, and into the UVA, UVB, and UVC regions, hence providing full coverage for dissipation of the past and present incident surface solar spectra (Michaelian, 2012; Michaelian and Simeonov, 2015). 
In contradiction to the thermodynamic dissipation theory, in the biological literature this phenomenon of full spectral coverage has been explained primarily through invoking the conventional wisdom of photoprotection (Demmig-Adams and Adams, 1992; Mulkidjanian and Junge, 1997; Wynn-Williams et al., 2002; Mulkidjanian et al., 2003; Castenholz and Garcia-Pichel, 2012).     

Photoprotective roles have especially been attributed to UV-absorbing biological pigments (e.g., mycosporine-like amino acids and scytonemins in cyanobacteria and algae; flavonoids, anthocyanins and phenylpropanoids in plants, etc.) since they don't seem to contribute to photosynthesis at all (Moisan and Mitchell, 2001). These theories usually trace the photoprotective role of UV-pigments back to the beginnings of cellular life in the early Archean when UV radiation was far more important component of the surface solar spectrum than it is today (Sagan, 1973; Mulkidjanian and Junge, 1997; Garcia-Pichel, 1998; Cockell and Knowland, 1999; Mulkidjanian et al., 2003). UV-screening ostensibly conferred pigment-containing organisms Darwinian advantages for survival in the harsh Archean environment of intense UV radiation. 

However, from a thermodynamic viewpoint, the UV is the region of the solar spectrum with an energy per photon sufficient for directly reconfiguring covalent bonds and it also corresponds to the greatest possible entropy production potential per unit photon dissipated. Therefore, under the high UV ambient conditions of our primitive planet, non-equilibrium thermodynamic principles of increasing the entropy production of Earth in its solar environment were probably the motive force for the dissipative synthesis of these organic pigments (Michaelian, 2017). The evidence for this inextricable link between UV light and nascent life has been reinforced with the verifications for biogenicity of $\sim$ 3.5 Ga old euphotic stromatolitic formations (Walter et al., 1980; Awramik et al., 1983; Schopf and Packer, 1987; Schopf, 1993; Schopf et al., 2002; Tice and Lowe, 2004; Van Kranendonk et al., 2008) of evidently photosynthetically active, yet UVC bathed, microbial colonies of cyanobacteria-like organisms (Westall et al., 2006; Westall, 2009).

In this paper we discuss how the two major classes of cyanobacterial UV-absorbing pigments, the mycosporine-like amino acids and the scytonemins, whose occurrence in organisms today is regarded as a relic from Archean times (Garcia-Pichel, 1998), perfectly match the type of microscopic dissipative structure which non-equilibrium thermodynamic principles would predict for microscopic dissipative structuring under the Archean Earth conditions.

In the following section we separately detail a set of germane properties of both pigment classes, taken from the literature, whereas in the third section we demonstrate how these properties are consistent with several postulates made concerning photon dissipative structures in previous work (Michaelian and Simeonov, 2015). Also, in the third section, we compare the pigments' optical properties with our previous construct of the most probable Earth surface solar spectrum as a function of time (see Michaelian and Simeonov, 2015).

\clearpage

\section{Properties of cyanobacterial UV-absorbing pigments}

\subsection{Mycosporine-like amino acids (MAAs)}
MAAs and a related group of organic compounds called mycosporines represent a large family of colorless, low-molecular-weight ($<$ 400 u), water-soluble, usually intracellular secondary metabolites widespread in the biological world (Dunlap and Chalker, 1986; Carreto et al., 1990; Rosic and Dove, 2011). The exact number of compounds within this family is yet to be determined, since they have only relatively recently been discovered (for a historical overview, see Schick and Dunlap, 2002 and \u{R}ezanka et al., 2004), and novel molecular species are constantly being uncovered. Thus far, however, their number is around 40 (Wada et al., 2015). The name ``mycosporine'' has to do with these being originally isolated and chemically identified from mycelia of sporulating fungi, where it was thought they played a role in light-induced sporulation (Leach, 1965; Trione and Leach, 1969).

\subsubsection{Physicochemical properties}
Chemically both MAAs and mycosporines are alicyclic compounds (see Fig. 1) sharing a central 5-hydroxy-5-hydroxymethyl-2-methoxycyclohex-2-ene ring with an amino compound substituted at the third C-atom and either an oxo or an imino functionality at the first C-atom (Favre-Bonvin et al., 1976; Ito and Hirata, 1977; Arpin et al., 1979; Karentz, 2001). While most authors don't make a clear chemical distinction between the two groups, several authors (for example: Bandaranayake, 1998; Schick and Dunlap, 2002; Carreto et al., 2011; Molin\'{e} et al., 2014; Miyamoto et al., 2014) when using the term mycosporines refer only to those molecular species with a central amino-cyclohexenone chromophore (also called oxo-mycosporines), and when using the term MAAs refer only to molecules with a central amino-cyclohexenimine chromophore (also called imino-mycosporines). The N-substitution on C-3 with different amino acids or amino alcohols is what gives the diversity of molecular structures within both groups (Korbee et al., 2006; Sinha et al., 2007; Carreto and Carignan, 2011). Within the MAA group, the most common amino acid on the C-3 position is glycine, whereas they also have a second amino acid, amino alcohol or an enaminone system attached at the C-1 position (D'Agostino et al., 2016).

\begin{figure*}
\includegraphics[scale=0.6]{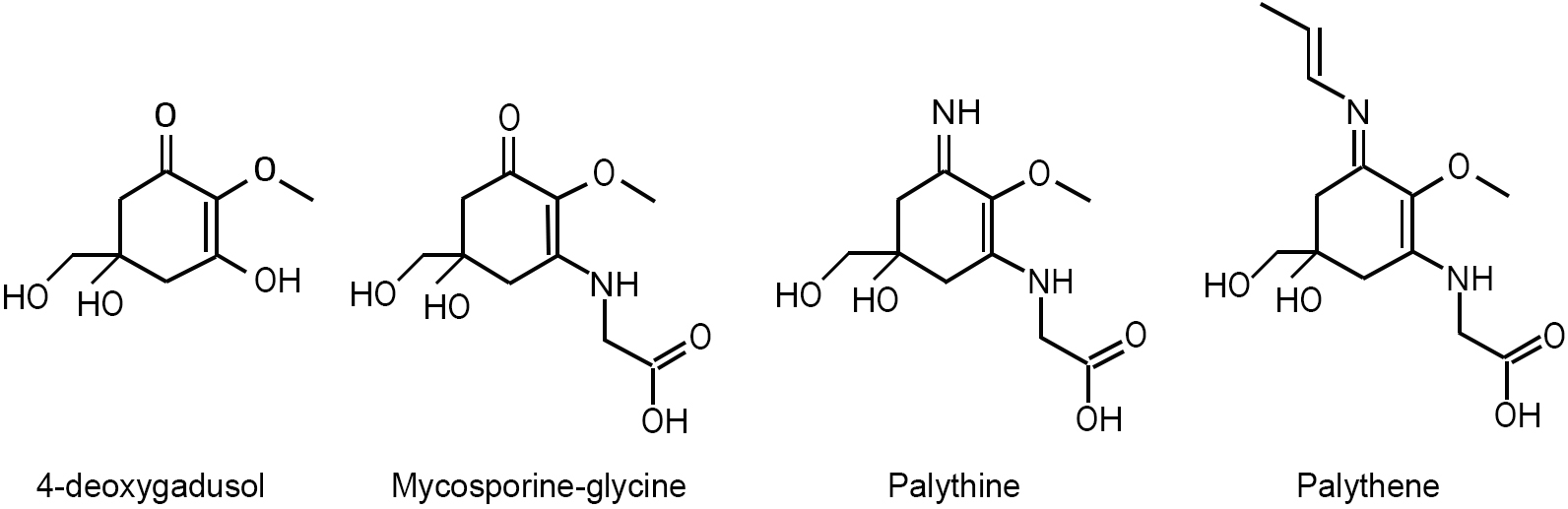}
\caption{Chemical structures of some common mycosporines and MAAs.}
\end{figure*}

This unique molecular structuring and bonding underlies their unique spectral properties. MAAs are one of the strongest UVA/UVB-absorbing substances in nature (Schmid et al., 2000); with wavelength absorption maxima $\lambda_{max}$ in the 310-362 nm interval and molar extinction coefficients $\varepsilon$ between 28,100 and 50,000 $M^{-1}cm^{-1}$ (Dunlap and Schick, 1998; Carreto et al., 2005; Gao and Garcia-Pichel, 2011). 
Their absorption spectra are characterized by a single sharp $\lambda_{max}$ with a bandwidth of approximately 20 nm and only about 2-3 nm apart from the $\lambda_{max}$ of structurally similar MAAs (see Fig. 2) which makes it very difficult to distinguish these compounds based solely on their absorption spectra (Karentz, 1994; Carroll and Shick, 1996). 

The values of $\lambda_{max}$ and $\varepsilon$ are dependent on the degree of derivatization of the central ring and the nature of the nitrogenous side groups (in particular the presence of additional conjugated double bonds) (Singh et al., 2010; Wada et al., 2015). 
Smaller, mono-substituted mycosporines (typically oxo-mycosporines) have their $\lambda_{max}$ values shifted to shorter wavelengths in the UVB and usually have lower $\varepsilon$ values; whereas MAAs (imino-mycosporines) are normally bi-substituted, with higher $\varepsilon$ values and $\lambda_{max}$ values shifted to longer wavelengths in the UVA (Portwich and Garcia-Pichel, 2003). 

For example, the direct metabolic precursor of all mycosporines, 4-deoxygadusol (Fig. 1), which has the minimal level of derivatization, has $\lambda_{max} = 268$ nm at acidic pH, and $\lambda_{max} = 294$ nm at basic pH; mycosporine-glycine (Fig. 1), the simplest oxo-mycosporine and direct precursor of all other mycosporines and MAAs has a $\lambda_{max} = 310$ nm, whereas palythine (Fig. 1), a simple mono-substituted MAA (amino-MAA), has $\lambda_{max} = 320$ nm and $\varepsilon = 36,200$ $M^{-1}cm^{-1}$  (Carreto et al., 2005; Gao and Garcia-Pichel, 2011). Palythene (Fig. 1), a bi-substituted MAA with an additional conjugated double bond, has one of the most red-shifted bands of all known MAA species, with $\lambda_{max} = 360$ nm and $\varepsilon = 50,000$ $M^{-1}cm^{-1}$ (Uemura et al., 1980).

The observed red-shift in $\lambda_{max}$ is a consequence of the degree of resonance delocalization inside the molecules; the more efficient is the electron delocalization (i.e. the stronger the $\pi$-conjugation character) the lower is the energy requirement for an electronic transition and consequently the higher are the $\lambda_{max}$ and $\varepsilon$ values (Carreto and Carignan, 2011; Wada et al., 2015).\\

\begin{figure*}
\includegraphics[scale=0.8]{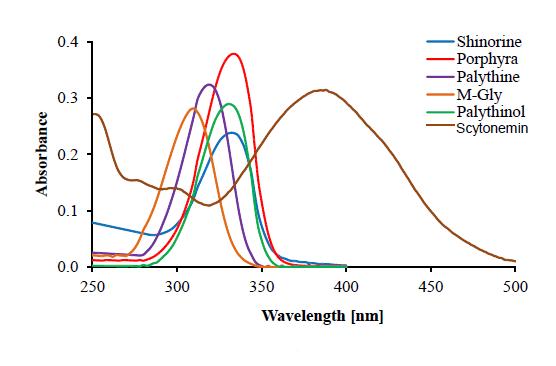}
\caption{Complementary absorption spectra of scytonemin and MAAs. (Adapted from Rastogi and Madamwar, 2016).}
\end{figure*} 

From a thermodynamic perspective, the fate of the electronic excitation energy is also a very relevant aspect of the absorption event since it is directly linked to the amount of entropy produced by the dissipative microscopic structure (i.e. pigment). Nonradiative, vibrational relaxation pathways of the excited state lead to more efficient energy dissipation and higher entropy production when compared to the fluorescent or phosphorescent radiative decay channels (W\"{u}rfel and Ruppel, 1985; Michaelian, 2011; 2012; 2016). In this respect MAAs prove to be very efficient dissipative structures, although all studies hitherto have discussed these thermodynamically relevant characteristics only from the standpoint of photostability and UV-photoprotection.

Aiming at fully describing their photophysical and photochemical properties and expanding the evidence on the assigned UV-photoprotective role, Inoue et al. (2002) and Conde et al. (2000; 2004; 2007) made several in vitro studies on the excited-state properties and photostability of various MAAs in aqueous solution (see Table 1). The results showed picoseconds excited state lifetimes, very low fluorescence quantum yields (e.g., $\phi_F$ (porphyra 334) $= 0.0016$), very low triplet formation quantum yields (e.g., $\phi_T$ (porphyra 334) $< 0.05$), and very low photodegradation quantum yields (e.g., $\phi_R$ (porphyra 334) = $2 - 4 \times 10^{-4}$) for all of the MAAs studied. These results are consistent with a very fast internal conversion (IC) process as the main deactivation pathway of the excited states, which was directly quantified by photoacoustic calorimetry experiments showing that $\sim$ 97 \% of the absorbed photonic energy is promptly dissipated into the surrounding medium as heat (Conde et al., 2004).

A computational study by Sampedro (2011) employing palythine as a model compound, confirmed these findings. The study indicates that the fast IC processes connecting the $S_2$/$S_1$ and $S_1$/$S_0$ states are due to the presence of two energetically accessible conical intersection points that can be reached by small geometrical changes in the atomic coordinates. It is now well established that conical intersections (a.k.a. molecular funnels or diabolic points) play a very important role in fast, non-radiative de-excitation transitions from excited electronic states to ground electronic state of molecules, particularly in many fundamental biological molecules, such as DNA/RNA, amino acids and peptides (Schermann, 2008). They enable effective coupling of the electronic degrees of freedom of the molecule to its phonon degrees of freedom, thereby facilitating radiationless decay by vibrational cooling to the ground state (in the process converting the absorbed high frequency UV photon into many low frequency infrared photons), which could make them examples of microscopic dissipative structuring of material in response to the impressed photon potential (Michaelian, 2016; 2017).

\subsubsection{Ecological distribution}
MAAs and mycosporines are cosmopolitan substances in ``optical'' habitats - planktonic, benthic and terrestrial; with the largest concentrations detected in environments exposed to high levels of solar irradiance (Castenholz and Garcia-Pichel, 2012 and references therein). They are now known to be the most common type of UV-absorbing natural substances, especially among aquatic organisms (Rastogi et al., 2010). 

While mycosporines have been reported only in the kingdom Fungi (mycosporine-glycine and mycosporine-taurine are exceptions), MAAs are more extensively distributed among taxonomically diverse organisms (Karsten, 2008; Carreto and Carignan, 2011). These include: cyanobacteria; heterotrophic bacteria; dinoflagellates, diatoms and other protists; red algae; green algae; various marine animals, especially corals and their associated biota (for a database on the distribution of MAAs, see Sinha et al., 2007). They seem to be completely absent from terrestrial plants, but are regularly found in terrestrial cyanobacteria (Garcia-Pichel and Castenholz, 1993) and terrestrial algae (Karsten et al., 2007).   

An interesting discovery by Ingalls et al. (2010) reveals that MAAs represent a considerable portion of the organic matter bound to diatom frustules, accounting for 3-27\% of the total carbon and 2-18\% of total nitrogen content of the frustules. Previously established views held that MAAs have mainly an intracellular location in these organisms.

\subsubsection{Biosynthesis}
The cyclohexenone core of MAAs is derived from intermediates of two fundamental anabolic pathways; the shikimate pathway (Favre-Bonvin et al., 1987; Shick et al., 1999; Portwich and Garcia-Pichel, 2003) and the pentose phosphate pathway (Balskus and Walsh, 2010), with the shikimate pathway being the predominant route for UV-induced MAA biosynthesis (Pope et al., 2015).
 
In MAA/mycosporine biosynthesis both pathways converge at a point where their respective 6-membered cyclic intermediates with similar structures are converted to 4-deoxygadusol (Fig. 1), the common precursor of all MAAs and mycosporines, a reaction catalyzed by the key enzyme O-methyltransferase (Pope et al., 2015; D'Agostino et al., 2016). 
    
These basic biochemical pathways lie at the heart of carbon metabolism, shared by all three domains of life; the shikimate pathway links carbohydrate catabolism to the biosynthesis of the aromatic amino acids and other aromatic biomolecules; similarly the pentose phosphate pathway uses glycolysis for the synthesis of pentose sugars, the nucleotide building blocks (Cohen, 2014). Thus, they are considered to have an ancient evolutionary origin, possibly even dating back to prebiotic times (Richards et al., 2006; Keller et al., 2014).

As mentioned in the previous section, a very interesting trait of MAAs is that they are extremely prevalent natural compounds produced by a variety of taxonomically very distant organisms from simple bacteria to algae and animals. A natural question arises: how can evolutionary so distant organisms share the same MAA encoding genes?

Several lines of evidence now suggest that the progenitor of the enzymatic machinery for MAA biosynthesis was probably a cyanobacterium or the cyanobacterial ancestor, while endosymbiotic events and prokaryote-to-eukaryote lateral gene transfer events during evolution account for their prevalence among all other biological taxa (Rozema et al., 2002; Waller et al., 2006; Starcevic et al., 2008; Singh et al., 2010; Singh et al., 2012).

\subsubsection{Function: traditional view vs. thermodynamic view}
Since their discovery in the 1960's, authors have struggled to confer specific physiological functions to MAAs. A UV-photoprotective role seemed most conspicuous, largely because of their unique UV-dissipating traits and the fact that their production is stimulated by UVB light. Later, this theory faced serious challenges, for example, the failure to find a correlation between intracellular MAA accumulation and UV-resistance in certain coral zooxanthellae (Kinzie, 1993), the phytoplankton \textit{Phaeocystis antarctica} (Karentz and Spero, 1995), the dinoflagellate \textit{Prorocentrum micans} (Lesser, 1996), certain cyanobacterial strains (Quesada and Vincent, 1997), and the red alga \textit{Palmaria palmata} (Karsten et al., 2003), etc.  

As a response, many researchers in the field came up with their own suggestions for MAA physiological roles, sometimes very different from the sunscreen role, such as; osmotic regulation, antioxidants, nitrogen storage, accessory pigments, protection from desiccation, protection from thermal stress, reproductive functions in fungi and marine invertebrates, etc.; all of which have also been challenged or discredited (for reviews of the different theories of MAA functions and the challenges they face, see: Korbee et al., 2006; Oren and Gunde-Cimerman, 2007; Rosic and Dove, 2011). 

From a traditional biological standpoint this apparent lack of a clear defining physiological function for these pigments looks extremely perplexing, especially when taking into account the extraordinary prevalence of these compounds in nature. Darwinian theory in its strictest traditional formulation, with evolution through natural selection operating only at the level of the individual, categorically dismisses this kind of phenomena; where an organism wastefully spends free energy and resources for the synthesis of metabolically expensive, nitrogen-containing compounds with no vital physiological function commensurate with their ubiquity and hence no, or little, benefit for its survival and reproduction. According to Darwinian theory, such a biosynthetic pathway, with little or no direct utility to the organism, should have been suppressed or completely eliminated through natural selection. However, exactly the opposite has happened in the course of evolution; MAA biosynthetic genes have not only survived but have undergone extensive dissemination across numerous taxa through horizontal gene transfer.

We postulate that the failure of Darwinian theory to allocate a niche for MAAs in its classical ``struggle for survival'' paradigm may be related to the fact that this theory is not founded on thermodynamic principles (for a discussion on this topic, see: Michaelian, 2011; 2012; 2016). 
From the perspective of non-equilibrium thermodynamics, MAAs do have a function and it is a thermodynamic function of energy dissipation, or, more generally, entropy production. This thermodynamic function can be readily inferred from their physicochemical properties related to photon dissipation described above. MAAs can be regarded as typical examples of microscopic dissipative structuring of carbon-based material for the overriding purpose of entropy production through highly efficient dissipation of high-frequency UV photons into heat (Michaelian, 2016; 2017). This may be the reason for the origin and proliferation over the surface of the Earth of all organic pigments and the reason for the ubiquity of organic pigments in the neighborhood of stars throughout the universe (Michaelian and Simeonov, 2017; Michaelian, 2016). 

This irreversible process of photon dissipation that MAAs and other bio-pigments perform, then couples to a secondary abiotic irreversible process of water evaporation from surfaces through the heat it releases into its aquatic milieu (Michaelian, 2012). Evidence exist that the profusion of life and chromophoric dissolved organic matter (CDOM) in the sea-surface microlayer (SML) causes significant heating of the ocean surface fomenting evaporation (Morel, 1988; Kahru et al., 1993; Jones et al., 2005; Patara et al., 2012) and even the irreversible process of hurricane formation and steering (Gnanadesikan et al., 2010). 

CDOM is the fraction of dissolved organic matter in water (DOM) that interacts with solar radiation (Nelson and Siegel, 2013). Light energy absorption by CDOM at the surface of the ocean usually exceeds that of phytoplankton pigments; $54\pm15$ percent of the total light absorption at 440 nm and $> 70$ \% of the total light absorption at 300 nm is due to CDOM (Siegel et al., 2002; Babin et al., 2003; Bricaud et al., 2010; Organelli et al., 2014). It is a complex and extremely variable mixture of organic pigments such as pheopigments (Bricaud et al., 2010), metal-free porphyrins (R\"{o}ttgers and Koch, 2012), humic and fulvic acids (Carlson and Mayer, 1980; Galgani and Engel, 2016), aromatic amino acids (Yamashita and Tanoue, 2003) and MAAs (Whitehead and Vernet, 2000; Steinberg et al., 2004; Tilstone et al., 2010). While it was previously believed that CDOM in the open ocean is chiefly a byproduct of heterotrophic organisms recycling phytoplankton cell contents (Nelson et al., 1998), more recent observations (Romera-Castillo et al., 2010) suggest a large contribution from active plankton exudation. 

Active secretion of MAAs into the surrounding water during surface blooms was demonstrated for the colonial cyanobacterium \textit{Trichodesmium spp} (Subramaniam et al., 1999; Steinberg et al., 2004), for the dinoflagellate \textit{Lingulodinium polyedrum} (Vernet and Whitehead, 1996; Whitehead and Vernet, 2000) and for the dinoflagellate \textit{Prorocentrum micans} (Tilstone et al., 2010). Interestingly, Tilstone et al. (2010) found far greater MAA concentration in the sea-surface microlayer samples when compared to the near-surface (0-2 m), and subsurface (0-110 m) samples. Whitehead and Vernet (2000) also concluded that free-floating MAAs contributed up to 10 \% of the UV absorption of the total DOM pool at 330 nm during the \textit{L. polyedrum} bloom.  This exudation of pigments by organisms into their environment would also seem to have little Darwinian advantage.  

All of the evidence presented suffices to conclude, with some certainty, that MAAs join in function most of the other bio-pigments in nature which act as catalysts for the dissipation of photons into heat at Earth's surface and the coupling of this heat to other abiotic entropy producing processes, such as; the water cycle, hurricanes, water and wind currents, etc.

\subsection{Scytonemins}
Scytonemin was first described in the 19th century by Swiss botanist Carl N\"{a}geli (N\"{a}geli, 1849; N\"{a}geli and Schwenderer, 1877) but was not isolated until 1991 when Garcia-Pichel and Castenholz (1991) made a more in-depth study of the compound. Proteau et al. (1993) determined the eight-ring, indolic-phenolic, homodimeric structure of scytonemin, which proved to be a completely novel structure unique among all hitherto known natural organic substances. For the carbon skeleton of this molecule they gave the trivial name ``the scytoneman skeleton''. Thus far, four additional substances with a scytoneman-type molecular structure have been isolated from cyanobacteria: dimethoxyscytonemin, tetramethoxyscytonemin, scytonine (Bultel-Ponc\'{e} et al., 2004) and scytonemin-imine (Grant and Louda, 2013); for which, in this review, we use the colloquial terms ``scytonemins'' or ``scytoneman pigments''.

\subsubsection{Physicochemical properties}
Scytonemin (Fig. 3), the representative and most common member of this yet poorly-explored family of aromatic indole alkaloids, is a relatively small molecule (544 u) built from two identical condensation products of tryptophanyl- and tyrosyl-derived subunits linked through a carbon-carbon bond (Proteau et al., 1993). Its IUPAC name is (3E,3'E)-3,3'-Bis(4-hydroxybenzylidene)-1,1'-bicyclopenta[b]indole-2,2'(3H,3'H)-dione.

Depending on the redox conditions it can exist in two inter-convertible forms: a predominant oxidized yellowish-brown form which is insoluble in water and only fairly soluble in organic solvents, such as pyridine and tetrahydrofuran, and a reduced form (Fig. 3) with bright red color that is slightly more soluble in organic solvents (Garcia-Pichel and Castenholz, 1991; Proteau et al., 1993). 
\begin{figure*}
\includegraphics[scale=0.6]{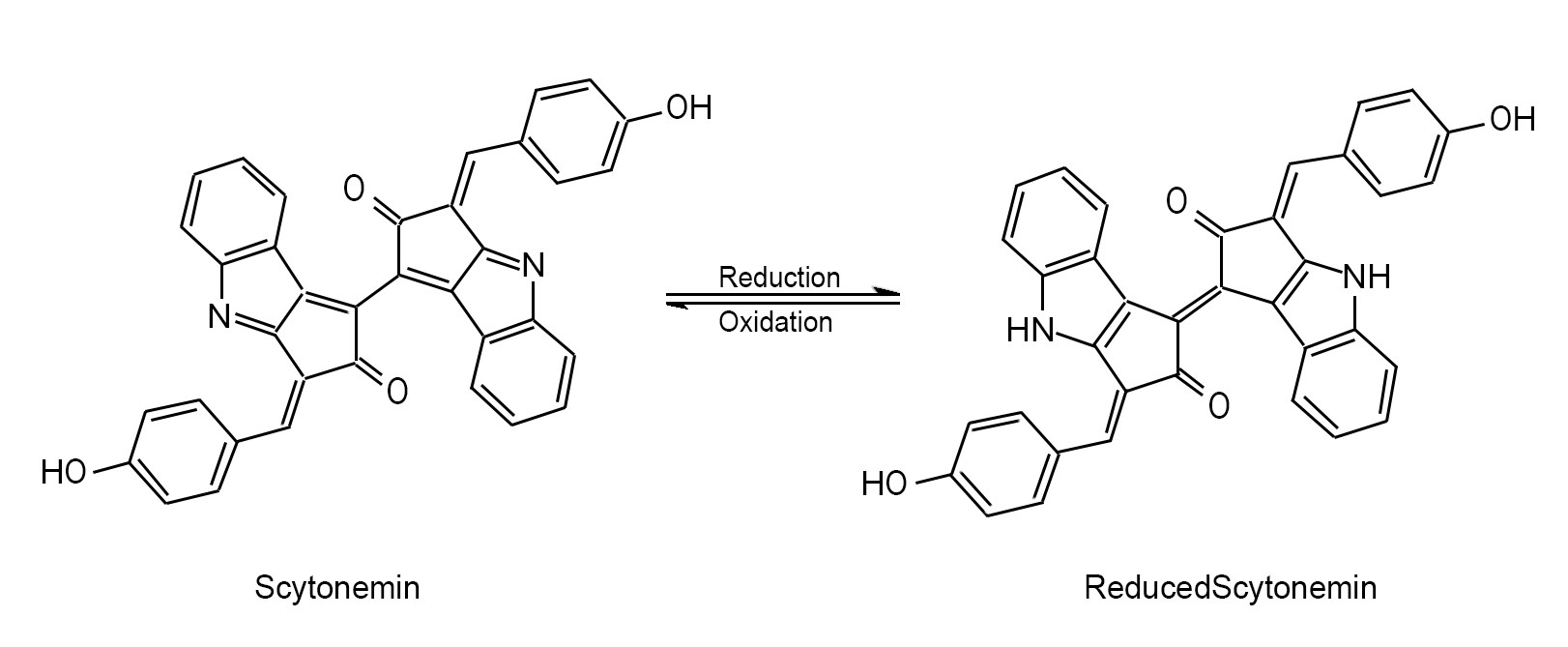}
\caption{Chemical structures of scytonemin and reduced scytonemin.}
\end{figure*}
Dimethoxy- and tetramethoxyscytonemin can be considered as derivatives of reduced scytonemin, where one or both of the ethenyl groups in the molecule have been saturated by two or four methoxy groups, respectively (Bultel-Ponc\'{e} et al., 2004; Varnali and Edwards, 2010). Another moderate degree of modification of the parent scytoneman skeleton can also be seen in scytonemin-3a-imine (a.k.a. scytonemin-imine), where the C-3a atom of scytonemin has been appended with a 2-imino-propyl radical (Grant and Louda, 2013).
 
Only the structure of scytonine deviates substantially from the dimeric scytoneman skeleton, due to the loss of one para-substituted phenol group and ring openings of both cyclopentenones where successive methoxylation and secondary cyclizations take place (Bultel-Ponc\'{e} et al., 2004). 

A full in-depth photophysical and photochemical characterization of scytonemins has yet to be performed; thus far only their elemental spectroscopic properties are known. 
Scytonemin absorbs very strongly and broadly across the UVC-UVB-UVA-violet-blue spectral region (see Fig. 2 and Fig. 4), with in vivo $\lambda_{max}$ at 370 nm and in vitro (tetrahydrofuran) $\lambda_{max}$ at 386 and 252 nm, with smaller peaks at 212, 278 and 300 nm (Garcia-Pichel and Castenholz, 1991; Garcia-Pichel et al., 1992; Sinha et al., 1999). Its observed long term persistence in cyanobacterial biocrusts or dried mats exposed to intense solar radiation might be an indication of exceptionally high photostability (Garcia-Pichel et al., 1992; Brenowitz and Castenholz, 1997; Fleming and Castenholz, 2007; Fulton et al., 2012; Lepot et al., 2014). 
 
Reduced scytonemin has a similar spectroscopic profile, with in vitro (tetrahydrofuran) $\lambda_{max}$ (nm) and $\varepsilon$ ($M^{-1}cm^{-1}$) values: 246 (30,000), 276 (14,000), 314 (15,000), 378 (22,000), 474 (14,000) and 572 (broad shoulder 7600) (Varnali and Edwards, 2014). 
A comparable absorption spectrum is also exhibited by scytonemin-imine, the mahogany-colored, polar derivative of scytonemin, with slightly different $\lambda_{max}$ values when measured in acetone (237, 366, 437 and 564 nm) and in ethanol (248, 305, 364, 440 and 553 nm) (Grant and Louda, 2013). 
\begin{figure*}
\includegraphics[scale=0.5]{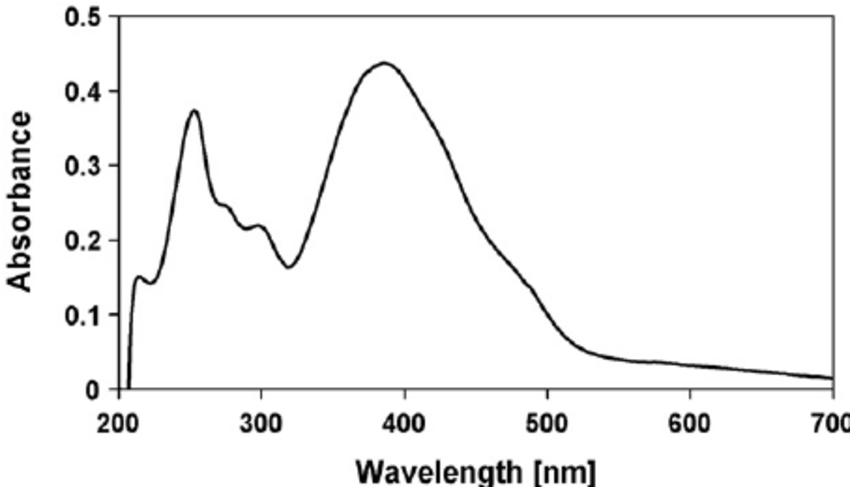}
\caption{In vitro absorption spectrum of scytonemin. (Adapted from Sinha et al., 1999).}
\end{figure*}
Contrary to these three scytoneman-type molecules, the methoxylated derivatives and scytonine do not absorb strongly in the UVA region but have high absorbances in the UVC, UVB and visible regions with in vitro (methanol) $\lambda_{max}$ (nm) and $\varepsilon$ ($M^{-1}cm^{-1}$) values for dimethoxyscytonemin: 316 (18,143) and 422 (23,015); for tetramethoxyscytonemin: 562 (5944); and for scytonine: 225 (37,054) and 270 (22,484) (Bultel-Ponc\'{e} et al., 2004). 

Concerning the monomeric scytoneman-type molecules nostodione A and prenostodione, isolated from natural cyanobacterial blooms, it remains debatable whether they are genuine cyanobacterial pigments or just intermediates in the biosynthesis of scytonemin (Ploutno and Carmeli, 2001; Soule et al., 2009a).

\subsubsection{Ecological distribution}
Unlike MAAs, scytonemins are exclusively cyanobacterial sheath pigments (Pathak et al., 2016). 
All phylogenetic lines of sheathed cyanobacteria contain scytonemins (Proteau et al., 1993), notably strains of the genera \textit{Nostoc}, \textit{Calothrix}, \textit{Scytonema}, \textit{Rivularia}, \textit{Chlorogloeopsis}, \textit{Lyngbya}, \textit{Hyella}, etc. (Sinha and H\"{a}der, 2008); as well as cyanolichens of the genera \textit{Peltula}, \textit{Collema} and \textit{Gonohymenia} (B\"{u}del et al., 1997). 

The mucilaginous extracellular sheath (matrix) consists of heteroglycans, peptides, proteins, DNA and different secondary metabolites (Pereira et al., 2009), where scytonemins are usually deposited in the outer layers, giving the sheath its distinctive dark yellow to brown color (Ehling-Schulz et al., 1997; Ehling-Schulz and Scherer, 1999). Up to 5\% of the dry weight of cultured scytonemin-synthesizing cyanobacteria is due to the pigment, but in natural assemblages this value can be even higher (Karsten et al., 1998). Curiously, Abed et al. (2010) reported two to six times higher concentrations of scytonemin than chlorophyll \textit{a} in cyanobacterial cryptobiotic soil crusts in the Oman Desert.  

Scytonemin-producing cyanobacteria typically inhabit highly insolated terrestrial, freshwater and coastal environments such as deserts, exposed rocks, cliffs, marine intertidal flats, oligotrophic lakes and ponds, ice shelves, hot springs, etc. (Castenholz and Garcia-Pichel, 2012 and references therein). In microbial mat communities, especially the extremophilic terrestrial and aquatic colonies, these cyanobacteria occupy the uppermost sunlit layers (Balskus et al., 2011). Scytonemin-imine, for example, was isolated from samples of natural \textit{Scytonema hoffmani} mats growing under high to intense (300-2000 $\mu$mol quanta $m^{-2}s^{-1}$) photon flux density (Grant and Louda, 2013). The methoxyscytonemins and scytonine were isolated alongside scytonemin from colonies of \textit{Scytonema sp.} growing on exposed granite at the Mitaraka inselberg in French Guyana, a region subjected to intense UVR-insolation (Bultel-Ponc\'{e} et al., 2004).

\subsubsection{Biosynthesis}
The biochemistry and genetics of cyanobacterial scytonemin biosynthesis has extensively been investigated by Soule et al. (2007; 2009a; 2009b), Balskus and Walsh (2008; 2009; 2011) and Sorrels et al. (2009). They have confirmed the assumption by Proteau et al. (1993), the discoverers of the scytonemin structure, that the scytoneman molecular scaffold is actually a condensation product of the aromatic amino acids tryptophan and tyrosine. Not only do these amino acids absorb in the UV themselves (Michaelian and Simeonov, 2015 and references therein), but they also serve as biosynthetic precursors for most known aromatic UV-absorbing bio-pigments, including: anthocyanins, flavonoids and phenylpropanoids in plants, melanins in heterotrophic organisms, scytonemins in cyanobacteria, etc. (Knaggs, 2003).

Sorrels et al. (2009) proposed an ancient evolutionary origin for the scytonemin biosynthetic pathway based on the combination of the fact that this gene cluster is highly conserved among evolutionary diverse strains of cyanobacteria (Soule et al., 2007; 2009a), and their own phylogenetic analyses implying that the cluster is under a purifying selection pressure. Intriguingly, Soule et al. (2009a) observed scytonemin biosynthetic genes even in some cyanobacterial strains incapable of producing the pigment (e.g., \textit{Anabaena} and \textit{Nodularia}), and interpreted this as a case of relic genetic information.

\subsubsection{Function: traditional view vs. thermodynamic view}
Similarly to MAAs, the Darwinian point of view can only describe scytonemin as an efficient protective biomolecule designed to filter out supposedly damaging high frequency UV radiation while at the same time allowing the transmittance of wavelengths necessary for photosynthesis (Ekebergh et al., 2015).

Within the framework of this traditional ``struggle for survival'' viewpoint, the majority of authors define scytonemins as an adaptive mechanism of extremophilic cyanobacteria that colonize harsh, inhospitable habitats experiencing high doses of UVR-insolation (Ehling-Schulz et al., 1997; Wynn-Williams et al., 1999; Hunsucker et al., 2001; Sinha and H\"{a}der, 2008; Rastogi et al., 2014). 

Among the evidence for the accredited photoprotective role is the discovery that up to 90\% of incident UV photons are prevented from entering sheathed, scytonemin-producing cyanobacterial cells, thus accomplishing significant reduction in chlorophyll \textit{a} photobleaching and maintaining photosynthetic efficiency (Garcia-Pichel and Castenholz, 1991; Garcia-Pichel et al., 1992). Other authors, in addition to the sunscreen role, ascribe supplementary defensive roles to scytonemin such as protection from oxidative, osmotic, heat and desiccation stress (Dillon et al., 2002; Matsui et al., 2012). 

Furthermore, scytonemin's superior UVC-absorbing capabilities in vivo, experimentally proven by treating cyanobacterial colonies with 0.5-1.0 $Wm^{-2}$ UVC radiation added to natural solar irradiance (Dillon and Castenholz, 1999), has led many authors to consider modern cyanobacterial production of scytonemins as a relic UV-protection mechanism from the pre-Great Oxygenation Event period (Garcia-Pichel, 1998; H\"{a}der et al., 2003).

Although it is beyond doubt that the efficient UV absorption and dissipation properties of the scytoneman pigments provide, to some extent, a beneficiary effect for the survival of sheathed cyanobacterial cells, the stance that this is the primary reason for the biological production of these pigments may be erroneous. Here are few examples of serious challenges and inconsistencies that the photoprotection paradigm faces:
\begin{enumerate}
\item Inability to explain the strong visible absorption bands of scytonemin-imine, tetramethoxyscytonemin and dimethoxyscytonemin where photosynthetic pigments absorb. The question is raised by Grant and Louda (2013): ``The absorption spectrum ($\lambda_{max}=$ 237, 366, 437, 564 nm in vitro), extending from the ultraviolet (UVB \& UVA) into the blue and green of the visible, appears to indicate a photoprotective role beyond shielding only UVR. That is, going on the premise that evolution generates and retains only advantageous secondary metabolites, then what is the role of the visible bands in this case?''
\item Inability to explain why many species of cyanobacteria do not synthesize scytonemins nor MAAs but, nevertheless, successfully cope with UV-induced cellular damage by employing only metabolic repair mechanisms (Quesada and Vincent, 1997; Castenholz and Garcia-Pichel, 2000).
\item Soule et al. (2007) developed scytoneminless mutant of the cyanobacterium \textit{Nostoc punctiforme} which proved to have indistinguishable growth rate from the wild type after both were subjected to UVA irradiation. The conclusion of the authors was that other photoprotective mechanisms can fully accommodate the absence of scytonemin in the mutant.
\end{enumerate}
In addition, very efficient absorption and dissipation of high-energy photons is not a prerequisite for photoprotection, but it is for dynamical dissipative structuring of material under an imposed photon potential. Nature has a simpler way of creating photostable molecules by making them either highly reflective or transparent to UV wavelengths (Michaelian, 2016).

These problems and paradoxes, generated when trying to explain scytonemins from within the Darwinian photoprotection paradigm, can be resolved by recurring to non-equilibrium thermodynamic principles. In this context, we will first address the question raised by Grant and Louda (2013) and then, based on all the evidence presented, we will assign a thermodynamic dissipative role to scytonemins. 

The seemingly paradoxical absorption spectra of scytonemin-imine and the methoxyscytonemins, which extend outside of the photoprotectively-relevant part of the spectrum, would make sense if these bio-pigments are understood as microscopic dissipative structures obeying non-equilibrium thermodynamic directive of increasing the global solar photon dissipation rate (Michaelian, 2013; Michaelian and Simeonov, 2015; Michaelian, 2016). 
Under this directive, one of the several ways to increase the global solar photon dissipation rate is by evolving (inventing) novel molecular structures (pigments) that cover ever more completely the prevailing surface solar spectrum (see Michaelian and Simeonov, 2015). 
This is precisely what is observed in the absorption spectra of the different scytoneman pigments. The strong visible absorption peaks of scytonemin-imine at 437 nm (violet) and 564 nm (green/yellow), of tetramethoxyscytonemin at 562 nm (green/yellow), and of dimethoxyscytonemin at 422 nm (violet) is exactly where the photosynthetic pigments do not peak in absorption (see, for example, Rowan, 1989). It is because of this rich assortment of diverse pigment molecules with complementary absorption bands that cyanobacterial biofilms, mats and soil crusts in nature tend to have high absorptivities, low albedos and appear almost black in color (Ustin et al., 2009). 

This fact leads us to an important conclusion on the thermodynamic function of the scytoneman pigments. We believe that it is most reasonable to consider the photon-dissipation role of scytonemins as the terrestrial analogue of the function that MAAs perform in the open aquatic environment. This assertion may be justified on their hydrophobic character and their inextricable connection to the extracellular polymeric substances (EPSs) of the cyanobacterial sheaths. Ekebergh et al. (2015) have shown that scytonemins have the greatest photostability in vivo, where they are embedded in their natural extracellular matrix milieu. These extracellular polymeric substances may therefore be playing the role of providing the dissipative medium required to disperse the excess vibrational energy after photon excitation of the pigment, bringing the system rapidly to the ground state and ready to absorb another photon.

In wet terrestrial regions of the planet, the thermodynamic role of photon dissipation coupled to the water cycle is performed mainly by the plant cover, but in arid and semi-arid lands, where vegetation is severely restricted, this function is allotted to microscopic assemblages of cyanobacteria, heterotrophic bacteria, algae and fungi known as biological soil crusts or biocrusts (Evans and Johansen, 1999; Belnap and Lange, 2001). It is theorized that these types of microbial communities represented life's pioneering on dry land and were the dominant ecosystem on the continents up until the colonization by land plants and animals in the Early Devonian (Beraldi-Campesi et al., 2014) brought on by the microbial retainment of water on the continents and the concomitant extension of the water cycle to regions far inland from the coasts. 

Michaelian (2013) postulated that bio-pigments are generally found in association with water because they use the high frequency vibrational modes of water molecules to facilitate their de-excitation. In this context we emphasize the fact that cyanobacteria isolated from dry regions display very high capacity to excrete large amounts of EPSs (Huang et al., 1998; Hu et al., 2003; Roeselers et al., 2007; Rossi et al., 2012), which are the main constituent of the biofilm matrix and together with microbial filaments play a key structural role in forming the biocrusts (Mager and Thomas, 2010; Karunakaran et al., 2011). The unique hydrophilic/hydrophobic nature of the EPSs enables highly efficient water capture and water storage within the biocrust by allowing the creation of moistened microenvironments where water dynamics is intricately regulated (Colica et al., 2014 and references therein). Hence, crust-covered soils are very hygroscopic and always exhibit higher water content compared to bare neighboring surfaces (Rossi and Phillips, 2015). 
This phenomenon is in accordance with our postulate that life's fundamental role is that of dispersing organic pigments and water over the entire surface of the Earth (Michaelian, 2013).     

A very conspicuous analogy between these terrestrial macroscopic and microscopic photon-dissipating biological ``carpets'' can be drawn. In the same manner as ecological succession of plant coverage leads to old climax forests with higher pigment content and lower albedos (Pokorny et al., 2010; Maes et al., 2011), ecological succession in biocrusts leads to increase in biomass of the late-stage scytonemin-producing cyanobacteria, and consequently accumulation of scytonemins in the matrix, an effect macroscopically observed as darkening of the biocrusted soil (i.e. decrease in albedo) (Couradeau et al., 2016). 
During dry periods in deserts when water availability is very limited, the heat generated from scytonemin's photon dissipation is expected to go predominantly into sensible heat of the biocrusts instead of into the latent heat of vaporization of water, and this is exactly what Couradeau et al. (2016) found when they measured $\sim$ $10^{\circ}$C higher temperature of biocrust-covered, dark soils in comparison to bare soils.

\clearpage

\section{Discussion}
In a previous work (Michaelian and Simeonov, 2015) we posited five basic tendencies that organic pigment evolution on Earth would have followed if the overriding function of life was to increase global photon dissipation; (1) increases in the photon absorption cross section with respect to the pigment physical size, (2) decreases in the electronic excited state lifetimes of the pigments, (3) quenching of the radiative de-excitation channels (e.g., fluorescence), (4) greater coverage of the surface solar spectrum, and (5) pigment proliferation and dispersion over an ever greater surface area of Earth.

To examine whether these five tendencies are corroborated within the evolutionary tendencies of MAAs and scytonemins we compare their properties to those of the aromatic amino acids (AAAs) (see Table 1). Our reason for choosing the AAAs is twofold; (1) they are considered to be among the earliest chromophoric organic molecules used by life with a prebiotic origin (Michaelian, 2011; Michaelian and Simeonov, 2015), and (2) since both MAAs and scytonemins are derived from intermediates of the shikimate pathway for AAA biosynthesis they most likely appeared later in evolution compared to the AAAs, probably when the biosynthetic machinery for the synthesis of the AAAs was already robust; an event that most likely long predated 3.4 Ga, considering that Busch et al. (2016) demonstrated that the ancestral tryptophan synthase of the last universal common ancestor (LUCA) was already a highly sophisticated enzyme at 3.4 Ga. 

Based on the data presented in previous sections on the spectral coverage and biological and geographical ubiquity of these pigments, we can state with a high degree of certainty that the fourth and fifth tendencies of pigment spectral versatility and pigment proliferation, respectively, are satisfied for the evolutionary histories of MAAs and scytonemins. 

\begin{table}[]
\centering
\caption{Comparison between the photophysical properties of the aromatic amino acids and representative compounds of the major classes of cyanobacterial UV-absorbing pigments}
\label{my-label}
\resizebox{\textwidth}{!}{%
\begin{tabular}{@{}ccccc@{}}
\toprule
\textbf{UV-absorbing bio-pigments} & \textbf{$\lambda_{max}$ (nm)} & \textbf{$\varepsilon$ ($M^{-1}cm^{-1}$)} & \textbf{Electronic excited state lifetime (ns)} & \textbf{Fluorescence quantum yield ($\phi_F$)} \\ \midrule
\multicolumn{5}{|c|}{\textbf{Aromatic amino acids}} \\ \midrule
\multicolumn{1}{|c|}{Phenylalanine (a)} & \multicolumn{1}{c|}{257} & \multicolumn{1}{c|}{195} & \multicolumn{1}{c|}{7.5} & \multicolumn{1}{c|}{0.024} \\ \midrule
\multicolumn{1}{|c|}{Tyrosine (a)} & \multicolumn{1}{c|}{274} & \multicolumn{1}{c|}{1405} & \multicolumn{1}{c|}{2.5} & \multicolumn{1}{c|}{0.14} \\ \midrule
\multicolumn{1}{|c|}{Tryptophan (a)} & \multicolumn{1}{c|}{278} & \multicolumn{1}{c|}{5579} & \multicolumn{1}{c|}{3.03} & \multicolumn{1}{c|}{0.13} \\ \midrule
\multicolumn{5}{|c|}{\textbf{Mycosporines and MAAs}} \\ \midrule
\multicolumn{1}{|c|}{Gadusol (b)} & \multicolumn{1}{c|}{269} & \multicolumn{1}{c|}{12400} & \multicolumn{1}{c|}{-} & \multicolumn{1}{c|}{non-fluorescent} \\ \midrule
\multicolumn{1}{|c|}{Mycosporine-$\gamma$-aminobutyric acid (c)} & \multicolumn{1}{c|}{310} & \multicolumn{1}{c|}{28900} & \multicolumn{1}{c|}{-} & \multicolumn{1}{c|}{-} \\ \midrule
\multicolumn{1}{|c|}{Mycosporine-glutamic acid (c)} & \multicolumn{1}{c|}{311} & \multicolumn{1}{c|}{20900} & \multicolumn{1}{c|}{-} & \multicolumn{1}{c|}{-} \\ \midrule
\multicolumn{1}{|c|}{Palythine (b, c)} & \multicolumn{1}{c|}{320} & \multicolumn{1}{c|}{36200} & \multicolumn{1}{c|}{-} & \multicolumn{1}{c|}{non-fluorescent} \\ \midrule
\multicolumn{1}{|c|}{Shinorine (b)} & \multicolumn{1}{c|}{333} & \multicolumn{1}{c|}{44700} & \multicolumn{1}{c|}{0.35} & \multicolumn{1}{c|}{0.002} \\ \midrule
\multicolumn{1}{|c|}{Porphyra-334 (b)} & \multicolumn{1}{c|}{334} & \multicolumn{1}{c|}{42300} & \multicolumn{1}{c|}{0.4} & \multicolumn{1}{c|}{0.0016} \\ \midrule
\multicolumn{1}{|c|}{Palythene (c)} & \multicolumn{1}{c|}{360} & \multicolumn{1}{c|}{50000} & \multicolumn{1}{c|}{-} & \multicolumn{1}{c|}{-} \\ \midrule
\multicolumn{5}{|c|}{\textbf{Scytonemins}} \\ \midrule
\multicolumn{1}{|c|}{Scytonemin (d)} & \multicolumn{1}{c|}{\begin{tabular}[c]{@{}c@{}}252, \\ 278, \\ 300, \\ 384\end{tabular}} & \multicolumn{1}{c|}{-} & \multicolumn{1}{c|}{-} & \multicolumn{1}{c|}{non-fluorescent} \\ \midrule
\multicolumn{1}{|c|}{Reduced Scytonemin (d)} & \multicolumn{1}{c|}{\begin{tabular}[c]{@{}c@{}}246, \\ 276, \\ 314, \\ 378, \\ 474, \\ 572\end{tabular}} & \multicolumn{1}{c|}{\begin{tabular}[c]{@{}c@{}}30000, \\ 14000, \\ 15000, \\ 22000, \\ 14000, \\ 7600\end{tabular}} & \multicolumn{1}{c|}{-} & \multicolumn{1}{c|}{-} \\ \midrule
\multicolumn{1}{|c|}{Scytonemin-imine (e)} & \multicolumn{1}{c|}{\begin{tabular}[c]{@{}c@{}}237, \\ 366, \\ 437, \\ 564\end{tabular}} & \multicolumn{1}{c|}{-} & \multicolumn{1}{c|}{-} & \multicolumn{1}{c|}{-} \\ \midrule
\multicolumn{1}{|c|}{Dimethoxyscytonemin (d)} & \multicolumn{1}{c|}{\begin{tabular}[c]{@{}c@{}}316, \\ 422\end{tabular}} & \multicolumn{1}{c|}{\begin{tabular}[c]{@{}c@{}}18143, \\ 23015\end{tabular}} & \multicolumn{1}{c|}{-} & \multicolumn{1}{c|}{-} \\ \midrule
\multicolumn{1}{|c|}{Tetramethoxyscytonemin (d)} & \multicolumn{1}{c|}{562} & \multicolumn{1}{c|}{5944} & \multicolumn{1}{c|}{-} & \multicolumn{1}{c|}{-} \\ \midrule
\multicolumn{1}{|c|}{Scytonine (d)} & \multicolumn{1}{c|}{\begin{tabular}[c]{@{}c@{}}225, \\ 270\end{tabular}} & \multicolumn{1}{c|}{\begin{tabular}[c]{@{}c@{}}37054, \\ 22484\end{tabular}} & \multicolumn{1}{c|}{-} & \multicolumn{1}{c|}{-} \\ \midrule
\multicolumn{5}{|c|}{\textbf{Other poorly characterized cyanobacterial UV-absorbing pigments}} \\ \midrule
\multicolumn{1}{|c|}{Gloeocapsin (f)} & \multicolumn{1}{c|}{392} & \multicolumn{1}{c|}{-} & \multicolumn{1}{c|}{-} & \multicolumn{1}{c|}{-} \\ \midrule
\multicolumn{1}{|c|}{Microcystbiopterins (g)} & \multicolumn{1}{c|}{\begin{tabular}[c]{@{}c@{}}$\sim$ 275, \\ $\sim$ 350\end{tabular}} & \multicolumn{1}{c|}{\begin{tabular}[c]{@{}c@{}}10000, \\ 3500\end{tabular}} & \multicolumn{1}{c|}{-} & \multicolumn{1}{c|}{-} \\ \midrule
\multicolumn{5}{l}{\begin{tiny} a)Berezin and Achilefu (2010); (b)Losantos et al. (2015a); (c)Wada et al. (2015); (d)Varnali and Edwards (2014); (e)Grant and Louda (2013); (f)Storme et al. (2015); (g)Lifshits et al. (2016).\end{tiny}}
\end{tabular}%
}
\end{table}

With respect to the first to third tendencies, in addition to the previously discussed material, we offer the data presented in Table 1. The data has been extracted from the available literature and as of 2019 is exhaustive. All of the compounds listed are representative members of their respective chemical groups. Gadusol is used instead of the more relevant compound 4-deoxygadusol because of lack of available data on 4-deoxygadusol and because of their chemical relatedness with similar spectroscopic properties (Losantos et al., 2015a, 2015b). The $\lambda_{max}$ and $\varepsilon$ values of gadusol in water are pH-dependent: 268 nm at pH $<$ 7 and 296 nm at pH $\geq$ 7 (Losantos et al., 2015a), and in Table 1 we use the values for acidic pH, considering that the Archean seawater was probably slightly acidic with pH $\sim$ 6.5 (Holland, 2003).    
Absorption peaks and extinction coefficients below about 220 nm, we believe, are due to the ionization of the molecules, a process which could destroy them. Photon dissipation is not through a conical intersection at these very short wavelengths and for this reason they are omitted from Table 1 and Fig. 5.

\begin{figure*}
\includegraphics[scale=1.3]{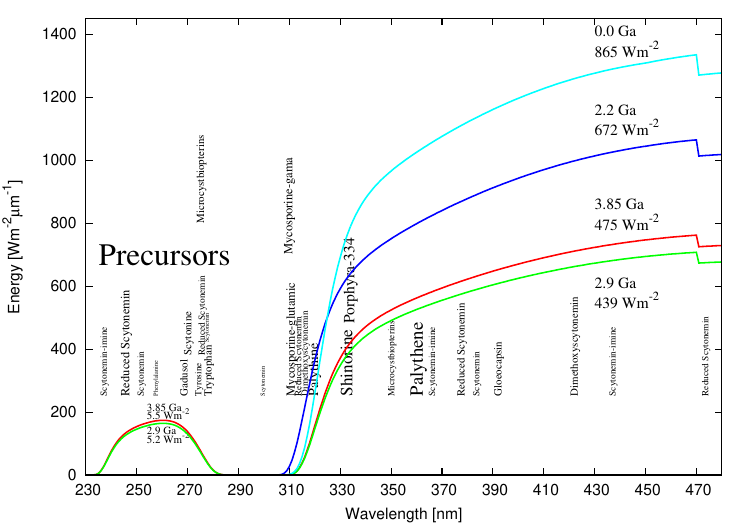}
\caption{The expected Earth surface solar spectrum at the given dates since present (Michaelian and Simeonov, 2015) and the maximum absorption ($\lambda_{max}$) of the mycosporine and scytonemin pigments and their aromatic amino acid precursors. The precursors have strong absorptions only in the UVC (230-280 nm) while scytonemins absorb strongly across the UVC, UVB, UVA, violet and blue regions, and mycosporines and MAAs usually have strong absorption in the UVB and UVA regions. Neither scytonemins nor MAAs peak strongly in the UVB region from $\sim$ 280-310 nm because aldehydes (\ce{CH_2O} and \ce{CH_3CHO}) produced by UVC light on common volcanic gases such as \ce{H_2S}, \ce{H_2O} and \ce{CO_2}, were absorbing strongly in this gap. This gap was later covered by \ce{O_2} and \ce{O_3} absorption after the Great Oxygenation Event at $\sim$ 2.3 Ga. The age of the given spectrum and the corresponding estimated integrated energy fluxes midday at the equator are listed.}
\end{figure*}

It is our hope that future experiments and studies into the nature and properties of these bio-pigments will help complete Table 1. However, even with the limited data available, presented in this article, a trend compatible with our conjecture is evident.  

Another conjecture made in Michaelian and Simeonov (2015) states that the surface solar spectrum wavelength region from approximately 280 to 310 nm has never reached the surface of the Earth during its entire geologic history; because during the Hadean and Archean eons these wavelengths were probably absorbed by atmospheric aldehydes (formaldehyde and acetaldehyde) (Sagan, 1973), and from the end of the Archean onwards gradual accumulation of oxygen and stratospheric ozone was responsible for their attenuation (Matsumi and Kawasaki, 2003; Stanley, 2008). In our earlier paper we also demonstrated how numerous fundamental molecules of life, common to all three domains of life, have strong absorbances across the long wavelength UVC, UVB and UVA regions except in this interval, which we used as an argument in favor of the thermodynamic dissipation theory of the origin and evolution of life (Michaelian, 2009; 2011; Michaelian, 2012).

Here we emphasize how this ``rule'' can also be applied to the cyanobacterial UV-absorbing pigments scytonemins and MAAs, which can be considered as evolutionary successors to the primordial pigments of life, specifically the AAAs. From the information presented in Table 1, Fig. 5, and Sect. 2, it is obvious that none of the compounds discovered so far, from both pigment groups, peak strongly in absorption inside this wavelength interval, which is consistent with our conjecture.
       
Indeed, the absorption spectrum of scytonemin (Fig. 4) has a very interesting shape which appears to follow the Archean surface spectrum. Although it is continuous from $\sim$ 220 to $\sim$ 700 nm, there is a dip in the $\sim$ 275 to $\sim$ 325 nm interval, and two large maxima at $\sim$ 250 nm and $\sim$ 380 nm. This is exactly the shape to be expected if the selective force for the evolution of this pigment was the dissipation of the Archean surface solar photon potential (Michaelian and Simeonov, 2015). Combining this crucial point with the previously discussed facts on scytonemin, it is tempting to speculate that this pigment had a key role in photon dissipation during the Archean, being capable of dissipating almost the entire Archean surface solar spectrum. The evolutionary invention of scytonemin's derivatives, as well as the mycosporines, the MAAs and still many other extinct and extant biological pigments, most likely resulted from the necessity to complement scytonemin's absorption with pigments that absorbed wavelengths reaching Earth's surface but were poorly absorbed by scytonemin itself. This kind of complementary spectral relationship between scytonemin and the MAAs has been well documented by several authors (e.g., Ehling-Schulz and Scherer, 1999; Ferroni et al., 2010; Castenholz and Garcia-Pichel, 2012) and is illustrated in Fig. 2. 

\clearpage

\section{Conclusions}
The available data on the ubiquity of pigments covering the region from the UVC to the infrared, many exuded by the organisms that produce them into the environment, make it increasingly difficult to assign to them a protective or antenna role within the Darwinian paradigm of the optimization of photosynthesis in benefit of the organism. We believe that sense can only be made of this by shifting the paradigm from one of ``photoprotection'' of the organism to the thermodynamically relevant paradigm of optimization of photon dissipation.

A number of contemporary pigment lines, most notably scytonemins and the mycosporine-like amino acids, appear to harbor relics of ancient biosynthetic production routes based on sugar and aromatic amino acid metabolism. These pigment lines absorb and dissipate efficiently in the UVC as well as the UVB, UVA and the visible. Many of these pigments are exuded into the environment which excludes the possibility of assigning them a role in photoprotection. Their strong absorption and dissipation in regions out of the photosynthetically active radiation (PAR) has been perplexing to perspectives within the Darwinian paradigm since the absorption by these pigments in these regions appears to have little utility to the organisms themselves. In fact, they absorb exactly where the photosynthetic pigments do not (and where water does not) and cover completely the Archean to present day Earth surface solar spectra.

It should be emphasized that our current knowledge of the diversity of cyanobacterial, algal and plant pigments and the thermodynamic function they perform is incomplete. For example, there are several indications of even richer diversity of UV-absorbing pigments in cyanobacteria than those hitherto characterized and classified into the two groups, mycosporines and scytonemins. The chemical structure and other elemental properties of one of these poorly investigated pigments, named gloeocapsin, have yet to be determined, but initial analysis suggests that it is chemically unrelated to both MAAs and scytonemins (Storme et al., 2015). Still other chemically distinct UV-absorbing cyanobacterial pigments, with a unique pterin structure, have been reported elsewhere (Matsunaga et al., 1993; Lifshits et al., 2016). The wavelengths of maximum absorption of these two ill-defined groups of cyanobacterial pigments are listed in Table 1 and are plotted in Fig. 5. As with the mycosporines and the scytonemins, their absorption properties are consistent with the optimization of dissipation of the prevailing photon spectrum at Earth's surface.

Taken as a whole, these data seem to indicate that, rather than photosynthesis being optimized under a Darwinian ``survival of the fittest'' paradigm, that the origin and evolution of life is driven by photon dissipation with the net effect of covering Earth's entire surface with pigments and water, reducing the albedo and the black-body temperature at which Earth radiates into space. It is our hope that this article will incite further investigation into the proposition that photon dissipation efficacy has been the fundamental driver of biological evolution on Earth.


\clearpage

\begin{thebibliography}{}

\bibitem[Abed et al.(2010)]{}
Abed, R. M. M., Al Kharusi, S., Schramm, A. and Robinson, M. D.: Bacterial diversity, pigments and nitrogen fixation of biological desert crusts from the Sultanate of Oman, FEMS Microbiol Ecol, 72, 418-428, 2010.

\bibitem[Arpin et al.(1979)]{LABEL}
Arpin, N., Curt, R. and Favre-Bonvin, J.: Mycosporines: mise au point et don\'{e}es nouvelles concernant leurs structures, leur distribution, leur localisation et leur biogen\`{e}se, Rev Mycol, 43, 247-257, 1979.

\bibitem[Awramik et al.(1983)]{LABEL}
Awramik, S. M., Schopf, J. W., and Walter, M. R.: Filamentous fossil bacteria from the Archean of Western Australia, Precambrian Res, 20, 357-374, 1983.

\bibitem[Babin et al.(2003)]{LABEL}
Babin, M., Stramski, D., Ferrari, G. M., Claustre, H., Bricaud, A., Obolensky, G., Hoepffner, N.: Variations in the light absorption coefficients of phyto-plankton, nonalgal particles, and dissolved organic matter in coastal waters around Europe, J. Geophys. Res, 108(C7), 3211, 2003. 

\bibitem[Balskus and Walsh (2008)]{LABEL}
Balskus, E. P. and Walsh, C. T.: Investigating the initial steps in the biosynthesis of cyanobacterial sunscreen scytonemin, J. Am. Chem. Soc, 130, 15260-15261, 2008. 

\bibitem[Balskus and Walsh (2009)]{LABEL}
Balskus, E. P., and Walsh, C. T.: An enzymatic cyclopentyl[b]indole formation involved in scytonemin biosynthesis, J. Am. Chem. Soc., 131, 14648-14649, 2009. 

\bibitem[Balskus and Walsh (2010)]{LABEL}
Balskus, E. P., and Walsh, C. T.: The genetic and molecular basis for sunscreen biosynthesis in cyanobacteria, Science, 329, 1653-1656, 2010.

\bibitem[Balskus et al.(2011)]{LABEL}
Balskus, E. P., Case, R. J. and Walsh, C. T.: The biosynthesis of cyanobacterial sunscreen scytonemin in intertidal microbial mat communities, FEMS Microbiol Ecol., 77(2), 322-332, 2011.

\bibitem[Bandaranayake (1998)]{LABEL}
Bandaranayake, W. M.: Mycosporines: are they nature's sunscreens?, Natural Product Reports, 1998, 159-171, 1998. 

\bibitem[Belnap and Lange (2001)]{LABEL}
Belnap, J. and Lange, O. L.: Biological Soil Crusts: Structure, Function, and Management, Springer Verlag, Berlin, Germany, 2001.

\bibitem[Beraldi-Campesi et al. (2014)]{LABEL}
Beraldi-Campesi, H., Farmer, J. D. and Garcia-Pichel, F.: Modern terrestrial sedimentary biostructures and their fossil analogs in mesoproterozoic subaerial deposits, Palaios, 29, 45-54, 2014.

\bibitem[Berezin and Achilefu (2010)]{LABEL}
Berezin, M. Y. and Achilefu, S.: Fluorescence Lifetime Measurements and Biological Imaging, Chem Rev., 110(5), 2641-2684, 2010. 

\bibitem[Brenowitz and Castenholz (1997)]{LABEL}
Brenowitz, S. and Castenholz, R. W.: Long-term effects of UV and visible irradiance on natural populations of a scytonemin-containing cyanobacterium (Calothrix sp.), FEMS Microbiol Ecol, 24, 343-352, 1997.

\bibitem[Bricaud et al. (2010)]{LABEL}
Bricaud, A., Babin, M., Claustre, H., Ras, J. and Tieche, F.: Light absorption properties and absorption budget of Southeast Pacific waters, J. Geophys. Res., 115, C08009, 2010.

\bibitem[B\"{u}del et al. (1997)]{LABEL}
B\"{u}del, B., Karsten, U. and Garcia-Pichel, F.: Ultraviolet absorbing scytonemin and mycosporine-like amino acid derivatives in exposed, rock-inhabiting cyanobacterial lichens, Oecologia, 112, 165-172, 1997.

\bibitem[Bultel-Ponc\'{e} et al. (2004)]{LABEL}
Bultel-Ponc\'{e}, V., Felix-Theodose, F., Sarlhou, C., Ponge, J. F. and Bodo, B.: New pigments from the terrestrial cyanobacterium Scytonema sp. collected on the Mitakara Inselberg, French Guyana, J Nat Prod, 67, 678-681, 2004.

\bibitem[Busch et al. (2016)]{LABEL}
Busch, F., Rajendran, C., Heyn, K., Schlee, S., Merkl, R. and Sterner, R.: Ancestral Tryptophan Synthase Reveals Functional Sophistication of Primordial Enzyme Complexes, Cell Chem Biol, 23(6), 709-715, 2016.

\bibitem[Carlson and Mayer (1980)]{LABEL}
Carlson, D. J. and Mayer, L. M.: Enrichment of dissolved phenolic material in the surface microlayer of coastal waters, Nature, 286, 482-483, 1980.

\bibitem[Carreto and Carignan (2011)]{LABEL}
Carreto, J. I. and Carignan, M. O.: Mycosporine-like amino acids: relevant secondary metabolites, Chemical and ecological aspects, Mar Drugs, 9, 387-446, 2011.

\bibitem[Carreto et al. (1990)]{LABEL}
Carreto, J. I., Carignan, M. O., Daleo, G. and De Marco, S. G.: Occurrence of mycosporine-like amino acids in the red-tide dinoflagellate Alexandrium excavatum: UV-photoprotective compounds?, J Plankton Res, 12, 909-921, 1990. 

\bibitem[Carreto et al. (2005)]{LABEL}
Carreto, J. I., Carignan, M. O. and Montoya, N. G.: A high-resolution reverse-phase liquid chromatography method for the analysis of mycosporine-like amino acids (MAAs) in marine organisms, Mar Biol, 146, 237-252, 2005.

\bibitem[Carreto et al. (2011)]{LABEL}
Carreto, J. I., Roy, S., Whitehead, K., Llewellyn, C. A. and Carignan, M. O.: UV-absorbing 'pigments': mycosporine-like amino acids. In: Roy, S., Llewellyn, C. A., Egeland, E. S. and Johnsen, G. (eds.) Phytoplankton pigments: characterisation, chemotaxonomy and applications in oceanography, Cambridge University Press, Cambridge, 412-441, 2011.

\bibitem[Carroll and Shick (1996)]{LABEL}
Carroll, A. K. and Shick, J. M.: Dietary accumulation of UV-absorbing mycosporine-like amino acids (MAAs) by the green sea urchin (Strongylocentrotus droebachiensis), Marine Biology, 124, 561-569, 1996.

\bibitem[Castenholz and Garcia-Pichel (2000)]{LABEL}
Castenholz, R. W. and Garcia-Pichel, F.: Cyanobacterial responses to UV-radiation. In: Whitton, B. A. and Potts, M. (eds.) The ecology of cyanobacteria. Their diversity in time and space, Kluwer Academic Publishers, Dordrecht, pp 591-611, 669, 2000. 

\bibitem[Castenholz and Garcia-Pichel (2012)]{LABEL}
Castenholz, R. W. and Garcia-Pichel, F.: Cyanobacterial responses to UV radiation, In: Whitton, B. A. (ed.) Ecology of Cyanobacteria II: Their Diversity in Space and Time, Springer Netherlands, 481-499, 2012. 

\bibitem[Cockell and Knowland (1999)]{LABEL}
Cockell, C. S. and Knowland, J.: Ultraviolet radiation screening compounds, Biol. Rev., 74, 311-345, 1999.

\bibitem[Cohen (2014)]{LABEL}
Cohen, G. N.: Microbial Biochemistry, (3rd edition), Springer, Netherlands, pages: 85-90 and 415-440, 2014. 

\bibitem[Colica et al. (2014)]{LABEL}
Colica, G., Li, H., Rossi, F., Li, D., Liu, Y. and de Philippis, R.: Microbial secreted exopolysaccharides affect the hydrological behavior of induced biological soil crusts in desert sandy soils, Soil Biol. Biochem., 68, 62-70, 2014.

\bibitem[Conde et al. (2000)]{LABEL}
Conde, F. R., Churio, M. S. and Previtali, C. M.: The photoprotector mechanism of mycosporine-like amino acids. Excited state properties and photostability of porphyra-334 in aqueous solution, J. Photochem. Photobiol. B, 56, 139-144, 2000.

\bibitem[Conde et al. (2004)]{LABEL}
Conde, F. R., Churio, M. S. and Previtali, C. M.: The deactivation pathways of the excited-states of the mycosporine-like amino acids shinorine and porphyra-334 in aqueous solution, Photochem. Photobiol. Sci., 3, 960-967, 2004.

\bibitem[Conde et al. (2007)]{LABEL}
Conde, F. R., Churio, M. S. and Previtali, C. M.: Experimental study of excited-state properties and photostability of the mycosporine-like amino acid palythine in water solution, Photochem. Photobiol. Sci., 6, 669-674, 2007.

\bibitem[Couradeau et al. (2016)]{LABEL}
Couradeau, E., Karaoz, U., Lim, H. C., Nunes da Rocha, U., Northern, T., Brodie, E. and Garcia-Pichel, F.: Bacteria increase arid-land soil surface temperature through the production of sunscreens, Nat. Commun., 7, 10373, 2016.

\bibitem[D'Agostino et al. (2016)]{LABEL}
D'Agostino, P. M., Javalkote, V. S., Mazmouz, R., Pickford, R., Puranik, P. R. and Neilan, B. A.: Comparative profiling and discovery of novel glycosylated mycosporine-like amino acids in two strains of the cyanobacterium Scytonema cf. crispum, Appl Environ Microbiol, 82, 5951-5959, 2016.

\bibitem[Demmig-Adams and Adams (1992)]{LABEL}
Demmig-Adams, B. and Adams, W. W.: Photoprotection and other responses of plants to high light stress, Annual Review of Plant Physiology and Plant Molecular Biology, 43, 599-626, 1992.

\bibitem[Dillon and Castenholz (1999)]{LABEL}
Dillon, J. G. and Castenholz, R. W.: Scytonemin, a cyanobacterial sheath pigment, protects against UVC radiation: implications for early photosynthetic life, J. Phycol., 35, 673-681, 1999.

\bibitem[Dillon et al. (2002)]{LABEL}
Dillon, J. G., Tatsumi, C. M., Tandingan, P. G. and Castenholz, R. W.: Effect of environmental factors on the synthesis of scytonemin, a UV-screening pigment, in a cyanobacterium (Chroococcidiopsis sp.), Arch. Microbiol., 177, 322-331, 2002.

\bibitem[Dunlap and Chalker (1986)]{LABEL}
Dunlap, W. C. and Chalker, B. E.: Identification and quantitation of near-UV absorbing compounds (S-320) in a hermatypic scleractinian, Coral Reefs, 5, 155-159, 1986.

\bibitem[Dunlap and Shick (1998)]{LABEL}
Dunlap, W. C. and Shick, J. M.: Ultraviolet radiation absorbing mycosporine-like amino acids in coral reef organisms: a biochemical and environmental environmental perspective, J Phycol, 34, 418-430, 1998.

\bibitem[Ehling-Schulz and Scherer (1999)]{LABEL}
Ehling-Schulz, M. and Scherer, S.: UV protection in cyanobacteria, Eur J Phycol, 34, 329-338, 1999.

\bibitem[Ehling-Schulz et al. (1997)]{LABEL}
Ehling-Schulz, M., Bilger, W. and Scherer, S.: UV-B-induced synthesis of photoprotective pigments and extracellular polysaccharides in the terrestrial cyanobacterium Nostoc commune, J Bacteriol., 179, 1940-1945, 1997.

\bibitem[Ekebergh et al. (2015)]{LABEL}
Ekebergh, A., Sandin, P. and M\.{a}rtensson, J.: On the photostability of scytonemin, analogues thereof and their monomeric counterparts, Photochem Photobiol Sci., 14(12), 2179-2186, 2015.

\bibitem[Evans and Johansen (1999)]{LABEL}
Evans, R. D. and Johansen, J. R.: Microbiotic crusts and ecosystem processes, Critical Reviews in Plant Sciences, 18, 183-225, 1999.

\bibitem[Favre-Bonvin et al. (1976)]{LABEL}
Favre-Bonvin, J., Arpin, N. and Brevard, C.: Structure de la mycosporine (P310), Can J Chem, 54, 1105-1113, 1976.

\bibitem[Favre-Bonvin et al. (1987)]{LABEL}
Favre-Bonvin, J., Bernillon, J., Salin, N. and Arpin, N.: Biosynthesis of mycosporines: Mycosporine glutaminol in Trichothecium roseum, Phytochemistry, 26, 2509-2514, 1987. 

\bibitem[Ferroni et al. (2010)]{LABEL}
Ferroni, L., Klisch, M., Pancaldi, S. and H\"{a}der, D. P.: Complementary UV-Absorption of Mycosporine-like Amino Acids and Scytonemin is Responsible for the UV-Insensitivity of Photosynthesis in Nostoc flagelliforme, Mar Drugs, 8(1), 106-121, 2010.

\bibitem[Fleming and Castenholz (2007)]{LABEL}
Fleming, E. D. and Castenholz, R. W.: Effects of periodic desiccation on the synthesis of the UV-screening compound, scytonemin, in cyanobacteria, Environ Microbiol, 9, 1448-1455, 2007.

\bibitem[Fulton et al. (2012)]{LABEL}
Fulton, J. M., Arthur, M. A. and Freeman, K. H.: Subboreal aridity and scytonemin in the Holocene Black Sea, Organic Geochemistry, 49, 47-55, 2012.

\bibitem[Galgani and Engel (2016)]{LABEL}
Galgani, L. and Engel, A.: Changes in optical characteristics of surface microlayers hint to photochemically and microbially mediated DOM turnover in the upwelling region off the coast of Peru, Biogeosciences, 13, 2453-2473, 2016.

\bibitem[Gao and Garcia-Pichel (2011)]{LABEL}
Gao, Q. and Garcia-Pichel, F.: Microbial ultraviolet sunscreens, Nat Rev Microbiol, 9, 791-802, 2011.

\bibitem[Garcia-Pichel (1998)]{LABEL}
Garcia-Pichel, F.: Solar ultraviolet and the evolutionary history of cyanobacteria, Origins Life Evol. B., 28, 321-347, 1998.

\bibitem[Garcia-Pichel and Castenholz (1991)]{LABEL}
Garcia-Pichel, F. and Castenholz, R. W.: Characterization and biological implications of scytonemin, a cyanobacterial sheath pigment, J Phycol, 27, 395-409, 1991.

\bibitem[Garcia-Pichel and Castenholz (1993)]{LABEL}
Garcia-Pichel, F. and Castenholz, R. W.: Occurrence of UV absorbing, mycosporine-like compounds among cyanobacterial isolates and an estimate of their screening capacity, Appl Environ Microbiol, 59, 163-169, 1993.

\bibitem[Garcia-Pichel et al. (1992)]{LABEL}
Garcia-Pichel, F., Sherry, N. D. and Castenholz, R. W.: Evidence for a UV sunscreen role of the extracellular pigment scytonemin in the terrestrial cyanobacterium Chlorogloeopsis sp., Photochem Photobiol, 56, 17-23, 1992.

\bibitem[Glansdorff and Prigogine (1971)]{LABEL}
Glansdorff, P. and Prigogine, I.: Thermodynamic Theory of Structure, Stability, and Fluctuations, Wiley-Interscience, London, 1971.

\bibitem[Gnanadesikan et al. (2010)]{LABEL}
Gnanadesikan, A., Emanuel, K., Vecchi, G. A., Anderson, W. G., and Hallberg, R.: How ocean color can steer Pacific tropical cyclones, Geophys. Res. Lett., 37, L18802, 2010.

\bibitem[Grant and Louda (2013)]{LABEL}
Grant, C. S. and Louda, J. W.: Scytonemin-imine, a mahogany-colored UV/Vis sunscreen of cyanobacteria exposed to intense solar radiation, Organic Geochemistry, 65, 29-36, 2013.

\bibitem[Gupta et al. (2015)]{LABEL}
Gupta, S., Guttman, M., Leverenz, R. L., Zhumadilova, K., Pawlowski, E. G., Petzold, C. J., Lee, K. K., Ralston, C. Y. and Kerfeld, C. A.: Local and global structural drivers for the photoactivation of the orange carotenoid protein, Proc Natl Acad Sci USA, 112(41), E5567-74, 2015.

\bibitem[H\"{a}der et al. (2003)]{LABEL}
H\"{a}der, D. P., Kumar, H. D., Smith, R. C. and Worrest, R. C.: Aquatic ecosystems: effects of solar ultraviolet radiation and interactions with other climatic change factors, Photochem. Photobiol. Sci., 2, 39-50, 2003.

\bibitem[Holland (2003)]{LABEL}
Holland, H. D.: The Geologic History of Seawater, In: Holland, H. D. and Turekian, K. K. (eds.), Treatise on Geochemistry, Volume 6; (ISBN: 0-08-044341-9); pp. 583-625, Elsevier Science, 2003. 

\bibitem[Horton et al. (1996)]{LABEL}
Horton, P., Ruban, A. V. and Walters, R. G.: Regulation of light harvesting in green plants, Ann. Rev. Plant Physiol. Plant Molec. Biol., 47, 655-684, 1996.

\bibitem[Hu et al. (2003)]{LABEL}
Hu, C., Zhang, D., Huang, Z. and Liu, Y.: The vertical microdistribution of cyanobacteria and green algae within desert crusts and the development of the algal crusts, Plant Soil., 257, 97-111, 2003.

\bibitem[Huang et al. (1998)]{LABEL}
Huang, Z., Liu, Y., Paulsen, B. S. and Klaveness, D.: Studies on Polysaccharides from Three Edible Species of Nostoc (Cyanobacteria) with Different Colony Morphologies: Comparison of Monosaccharide Compositions and Viscosities of Polysaccharides from Field Colonies and Suspension Cultures, J. Phycol., 34, 962-968, 1998.

\bibitem[Hunsucker et al. (2001)]{LABEL}
Hunsucker, S. W., Tissue, B. M., Potts, M. and Helm, R. F.: Screening protocol for the ultraviolet-photoprotective pigment scytonemin, Analyt. Biochem., 288, 227-230, 2001.

\bibitem[Inoue et al. (2002)]{LABEL}
Inoue, Y., Hori, H., Sakurai, T., Tokitomo, Y., Saito, J. and Misonou, T.: Measurement of Fluorescence Quantum Yield of Ultraviolet-Absorbing Substance Extracted from Red Alga: Porphyra yezoensis and its Photothermal Spectroscopy, Optical Review, 9(2), 75-80, 2002. https://doi.org/10.1007/s10043-002-0075-3

\bibitem[Ingalls et al. (2010)]{LABEL}
Ingalls, A. E., Whitehead, K. and Bridoux, M. C.: Tinted windows: The presence of the UV absorbing compounds called mycosporine-like amino acids embedded in the frustules of marine diatoms, Geochim. Cosmochim. Acta, 74, 104-115, 2010.

\bibitem[Ito and Hirata (1977)]{LABEL}
Ito, S. and Hirata, Y.: Isolation and structure of a mycosporine from the zoanthidian Palythoa tuberculosa, Tetrahedron Lett., 28, 2429-2430, 1977.

\bibitem[Jones et al. (2005)]{LABEL}
Jones, I., George, G. and Reynolds, C.: Quantifying effects of phytoplankton on the heat budgets of two large limnetic enclosures, Freshwater Biol., 50, 1239-1247, 2005.

\bibitem[Kahru et al. (1993)]{LABEL}
Kahru, M., Leppanen, J. M., and Rud, O.: Cyanobacterial blooms cause heating of the sea surface, Mar. Ecol.-Prog. Ser., 101, 1-7, 1993.

\bibitem[Karentz (1994)]{LABEL}
Karentz, D.: Ultraviolet tolerance mechanisms in Antarctic marine organisms, Antarctic Research Series, 62, 93-110, 1994.

\bibitem[Karentz (2001)]{LABEL}
Karentz, D.: Chemical defenses of marine organisms against solar radiation exposure: UV absorbing mycosporine-like amino acids and scytonemin. In: McClintock, J. B. and Baker, B. J. (eds.) Marine chemical ecology, CRC Press, Boca Raton, pp 481-519, 2001.

\bibitem[Karentz and Spero (1995)]{LABEL}
Karentz, D. and Spero, H. J.: Response of natural Phaeocystis population to ambient fluctuations of UVB radiation caused by Antarctic ozone depletion, Journal of Plankton Research, 17, 1771-1789, 1995.

\bibitem[Karsten (2008)]{LABEL}
Karsten, U.: Defense strategies of algae and cyanobacteria against solar ultraviolet radiation. In: Amsler, C. D. (ed.), Algal Chemical Ecology, Springer-Verlag, Berlin Heidelberg, 2008.

\bibitem[Karsten et al. (2003)]{LABEL}
Karsten, U., Dummermuth, A., Hoyer, K. and Wiencke, C.: Interactive effects of ultraviolet radiation and salinity on the ecophysiology of two Arctic red algae from shallow waters, Polar Biology, 26, 249-258, 2003.

\bibitem[Karsten et al. (2007)]{LABEL}
Karsten, U., Lembcke, S. and Schumann, R: The effect of ultraviolet radiation on photosynthetic performance, growth and sunscreen compound in aeroterrestrial biofilm algae isolated from building facades, Planta, 225, 991-1000, 2007.

\bibitem[Karsten et al. (1998)]{LABEL}
Karsten, U., Maier, J., and Garcia-Pichel, F.: Seasonality in UV absorbing compounds of cyanobacterial mat communities from an intertidal mangrove flat, Aquat. Microb. Ecol., 16, 37-44, 1998.

\bibitem[Karunakaran et al. (2011)]{LABEL}
Karunakaran, E., Mukherjee, J., Ramalingam, B. and Biggs, C. A.: "Biofilmology": a multidisciplinary review of the study of microbial biofilms, Appl. Microbiol. Biotechnol., 90, 1869-1881, 2011.

\bibitem[Keller et al. (2014)]{LABEL}
Keller, M. A., Turchyn, A. V. and Ralser, M.: Non-enzymatic glycolysis and pentose phosphate pathway-like reactions in a plausible Archean ocean, Mol Syst Biol., 10, 725, 2014.

\bibitem[Kinzie (1993)]{LABEL}
Kinzie, R. A. III: Effects of ambient levels of solar ultraviolet radiation on zooxanthellae and photosynthesis of the reef coral Montipora verrucosa, Marine Biology, 116, 319-327, 1993.

\bibitem[Knaggs (2003)]{LABEL}
Knaggs, A. R.: The biosynthesis of shikimate metabolites, Nat. Prod. Rep., 20, 119-136, 2003.

\bibitem[Korbee et al. (2006)]{LABEL}
Korbee, N., Figueroa, F. L. and Aguilera, J.: Accumulation of mycosporine-like amino acids (MAAs): biosynthesis, photocontrol and ecophysiological functions, Revista Chilena de Historia Natural, 79, 119-132, 2006.

\bibitem[Leach (1965)]{LABEL}
Leach, C. M.: Ultraviolet absorbing substances associated with light-induced sporulation in fungi, Can. J. Bot., 43, 185-200, 1965.

\bibitem[Lepot et al. (2014)]{LABEL}
Lepot, K., Deremiens, L., Namsaraev, Z., Comp\`{e}re, P., G\'{e}rard, E., Verleyen, E., Tavernier, I., Hodgson, D. A., Wilmotte, A., and Javaux, E. J.: Organo-mineral imprints in fossil cyanobacterial mats of an Antarctic lake, Geobiology, 12, 424-450, 2014.

\bibitem[Lesser (1996)]{LABEL}
Lesser, M. P.: Acclimation of phytoplankton to UV-B radiation: oxidative stress and photo inhibition of photosynthesis are not prevented by UV-absorbing compounds in the dinoflagellate Prorocentrum micans, Mar. Ecol. Prog. Ser., 132, 287-297, 1996.

\bibitem[Lifshits et al. (2016)]{LABEL}
Lifshits, M., Kovalerchik, D. and Carmeli, S.: Microcystbiopterins A-E, five O-methylated biopterin glycosides from two Microcystis spp. bloom biomasses, Phytochemistry, 123, 69-74, 2016.

\bibitem[Losantos et al. (2015a)]{LABEL}
Losantos, R., Churio, M. S. and Sampedro, D.: Computational Exploration of the Photoprotective Potential of Gadusol, ChemistryOpen, 4, 155-160, 2015a.

\bibitem[Losantos et al. (2015b)]{LABEL}
Losantos, R., Sampedro, D. and Churio, M. S.: Photochemistry and photophysics of mycosporine-like amino acids and gadusols, nature's ultraviolet screens, Pure Appl. Chem., 87(9-10), 979-996, 2015b.

\bibitem[Maes et al. (2011)]{LABEL}
Maes, W. H., Pashuysen, T., Trabucco, A., Veroustraete, F. and Muys, B.: Does energy dissipation increase with ecosystem succession? Testing the ecosystem exergy theory combining theoretical simulations and thermal remote sensing observations, Ecological Modelling, 222, 3917-3941, 2011.

\bibitem[Mager and Thomas (2010)]{LABEL}
Mager, D. M. and Thomas, A. D.: Carbohydrates in cyanobacterial soil crusts as a source of carbon in the southwest Kalahari, Botswana, Soil Biol. Biochem., 42, 313-318, 2010.

\bibitem[Matsui et al. (2012)]{LABEL}
Matsui, K., Nazifi, E., Hirai, Y., Wada, N., Matsugo, S. and Sakamoto, T.: The cyanobacterial UV-absorbing pigment scytonemin displays radical scavenging activity, J. Gen. Appl. Microbiol., 58, 137-144, 2012.

\bibitem[Matsumi and Kawasaki (2003)]{LABEL}
Matsumi, Y. and Kawasaki, M.: Photolysis of Atmospheric Ozone in the Ultraviolet Region, Chem. Rev., 103, 4767-4781, 2003.

\bibitem[Matsunaga et al. (1993)]{LABEL}
Matsunaga, T., Burgess, G., Yamada, N., Komatsu, K., Yoshida, S. and Wachi, Y.: An ultraviolet (UV-A) absorbing biopterin glucoside from the marine planktonic cyanobacterium Oscillatoria sp., Appl. Microbiol. Biotechnol., 39, 250-253, 1993.

\bibitem[Michaelian (2009)]{LABEL}
Michaelian, K.: Thermodynamic origin of life, http://arxiv.org/abs/0907.0042 (last access: 25 July 2015), 2009.

\bibitem[Michaelian (2011)]{LABEL}
Michaelian, K.: Thermodynamic dissipation theory for the origin of life, Earth Syst. Dynam., 2, 37-51, 2011.

\bibitem[Michaelian (2012)]{LABEL}
Michaelian, K.: HESS Opinions "Biological catalysis of the hydrological cycle: life's thermodynamic function", Hydrol. Earth Syst. Sci., 16, 2629-2645, 2012.

\bibitem[Michaelian (2013)]{LABEL}
Michaelian, K.: A non-linear irreversible thermodynamic perspective on organic pigment proliferation and biological evolution, J. Phys. Conf. Ser. 475, 012010, 2013.

\bibitem[Michaelian (2016)]{LABEL}
Michaelian, K.: Thermodynamic Dissipation Theory of the Origin and Evolution of Life: Salient characteristics of RNA and DNA and other fundamental molecules suggest an origin of life driven by UV-C light, Self-published. Printed by CreateSpace, Mexico City, ISBN: 978-1541317482, 2016.

\bibitem[Michaelian (2017)]{LABEL}
Michaelian, K.: Microscopic dissipative structuring and proliferation at the origin of life, Heliyon, Volume 3, Issue 10, October 2017, e00424, ISSN 2405-8440 

\bibitem[Michaelian and Simeonov (2015)]{LABEL}
Michaelian, K. and Simeonov, A.: Fundamental molecules of life are pigments which arose and co-evolved as a response to the thermodynamic imperative of dissipating the prevailing solar spectrum, Biogeosciences, 12, 4913-4937, 2015.

\bibitem[Michaelian and Simeonov (2017)]{LABEL}
Michaelian, K. and Simeonov, A.: Thermodynamic explanation for the cosmic ubiquity of organic pigments, Cornell ArXiv, arXiv:1608.08847 [astro-ph.EP], 2016, and Astrobiol Outreach 2017, 5:1 DOI: 10.4172/2332-2519.1000156.

\bibitem[Miyamoto et al. (2014)]{LABEL}
Miyamoto, K. T., Komatsu, M. and Ikeda, H.: Discovery of gene cluster for mycosporine-like amino acid biosynthesis from Actinomycetales microorganisms and production of a novel mycosporine-like amino acid by heterologous expression, Appl Environ Microbiol., 80(16), 5028-5036, 2014.

\bibitem[Moisan and Mitchell (2001)]{LABEL}
Moisan, T. A. and Mitchell, B. G.: UV absorption by mycosporine-like amino acids in Phaeocystis antarctica Karsten induced by photosynthetically available radiation, Mar. Biol., 138 (1), pp. 217-227, 2001.

\bibitem[Molin\'{e} et al. (2014)]{LABEL}
Molin\'{e}, M., Libkind, D., de Garcia, V. and Giraudo, M. R.: Production of Pigments and Photo-Protective Compounds by Cold-Adapted Yeasts, In: Buzzini, P. and Margesin, R. (eds.) Cold-adapted Yeasts: Biodiversity, Adaptation Strategies and Biotechnological Significance, Springer Berlin Heidelberg, pp 193-224, 2014.

\bibitem[Morel (1988)]{LABEL}
Morel, A.: Optical modeling of the upper ocean in relation to its biogenous matter content (case I waters), J Geophys Res, 93(C9), 10749-10768, 1988.

\bibitem[Mulkidjanian and Junge (1997)]{LABEL}
Mulkidjanian, A. Y. and Junge, W.: On the origin of photosynthesis as inferred from sequence analysis, Photosynth. Res., 51, 27-42, 1997.

\bibitem[Mulkidjanian et al. (2003)]{LABEL}
Mulkidjanian, A. Y., Cherepanov, D. A., and Galperin, M. Y.: Survival of the fittest before the beginning of life: selection of the first oligonucleotide-like polymers by UV light, BMC Evol. Biol., 3, 12, 2003.

\bibitem[N\"{a}geli (1849)]{LABEL}
N\"{a}geli, C.: Gattungen einzelliger Algen, physiologisch und systematisch bearbeitet, Neue Denkschrift Allg Schweiz Natur Ges, 10, 1-138, 1849.

\bibitem[N\"{a}geli and Schwenderer (1877)]{LABEL}
N\"{a}geli, C. and Schwenderer, S.: Das Mikroskop, 2nd edn. Willhelm Engelmann Verlag, Leipzig, 505 p., 1877.

\bibitem[Nelson and Siegel (2013)]{LABEL}
Nelson, N. B. and Siegel, D. A.: The global distribution and dynamics of chromophoric dissolved organic matter, Ann Rev Mar Sci., 5, 447-476, 2013.

\bibitem[Nelson et al. (1998)]{LABEL}
Nelson, N. B., Siegel, D. A. and Michaels, A. F.: Seasonal dynamics of colored dissolved material in the Sargasso Sea, Deep-Sea Res. I, 45, 931-957, 1998.

\bibitem[Oren and Gunde-Cimerman (2007)]{LABEL}
Oren, A. and Gunde-Cimerman, N.: Mycosporines and mycosporinelike amino acids: UV protectants or multipurpose secondary metabolites? FEMS Microbiol Lett, 269, 1-10, 2007.

\bibitem[Organelli et al. (2014)]{LABEL}
Organelli, E., Bricaud, A., Antoine, D. and Matsouka, A.: Seasonal dynamics of light absorption by chromophoric dissolved organic matter (CDOM) in the NW Mediterranean Sea (BOUSSOLE site). Deep-Sea Research Part I, 91, 72-85, 2014. 

\bibitem[Patara et al. (2012)]{LABEL}
Patara, L., Vichi, M., Masina, S., Fogli, P. G. and Manzini, E.: Global response to solar radiation absorbed by phytoplankton in a coupled climate model, Clim Dyn, 39(7-8), 1951-1968, 2012.

\bibitem[Pathak et al. (2016)]{LABEL}
Pathak, J., Rajneesh, Richa, Sonker, A.S., Kannaujiya, V. K. and Sinha, R. P.: Cyanobacterial extracellular polysaccharide sheath pigment, scytonemin: A novel multipurpose pharmacophore, In: Se-Kwon, K. (ed.) Marine Glycobiology: Principles and Applications, 1st edition, Taylor \& Francis Group, 6000 Broken Sound Parkway NW, Suite 300, Boca Raton, FL 33487-2742 CRC Press, Pages 323-337, 2016.

\bibitem[Pereira et al. (2009)]{LABEL}
Pereira, S., Zille, A., Micheletti, E., Moradas-Ferreira, P., de Philippis, R. and Tamagnini, P. Complexity of cyanobacterial exopolysaccharides: Composition, structures, inducing factors and putative genes involved in their biosynthesis and assembly. FEMS Microbiol. Rev. 33, 917-941, 2009.

\bibitem[Ploutno and Carmeli (2001)]{LABEL}
Ploutno, A. and Carmeli, S.: Prenostodione, a novel UV-absorbing metabolite from a natural bloom of the cyanobacterium Nostoc species, J Nat Prod, 64, 544-545, 2001. 

\bibitem[Pokorny et al. (2010)]{LABEL}
Pokorny, J., Brom, J., Cermak, J., Hesslerova, P., Huryna, H., Nadezhdina, N. and Rejskova, A.: Solar energy dissipation and temperature control by water and plants, Int. J. Water, 5, 4, 2010.

\bibitem[Pope et al. (2015)]{LABEL}
Pope, M. A., Spence, E., Seralvo, V., Gacesa, R., Heidelberger, S., Weston, A. J., Dunlap, W. C., Shick, J. M. and Long, P. F.: O-Methyltransferase is shared between the pentose phosphate and shikimate pathways and is essential for mycosporine-like amino acid biosynthesis in Anabaena variabilis ATCC 29413, Chembiochem, 16, 320-327, 2015.

\bibitem[Portwich and Garcia-Pichel (2003)]{LABEL}
Portwich, A. and Garcia-Pichel, F.: Biosynthetic pathway of mycosporines (mycosporine-like amino acids) in the cyanobacterium Chlorogloeopsis sp. strain PCC 6912, Phycologia, 42, 384-392, 2003.

\bibitem[Prigogine (1967)]{LABEL}
Prigogine, I.: Thermodynamics of Irreversible Processes, Wiley, New York, 1967.

\bibitem[Proteau et al. (1993)]{LABEL}
Proteau, P. J., Gerwick, W. H., Garcia-Pichel, F. and Castenholz, R.: The structure of scytonemin, an ultraviolet sunscreen pigment from the sheaths of cyanobacteria, Experientia, 49, 825-829, 1993.

\bibitem[Quesada and Vincent (1997)]{LABEL}
Quesada, A. and Vincent, W. F.: Strategies of adaptation by Antarctic cyanobacteria to ultraviolet radiation, Eur J Phycol, 32, 335-342, 1997.

\bibitem[Rastogi and Madamwar (2016)]{LABEL}
Rastogi, R. P. and Madamwar, D.: Cyanobacteria Synthesize their own UV-Sunscreens for Photoprotection, Bioenergetics, 5, 138, 2016.

\bibitem[Rastogi et al. (2010)]{LABEL}
Rastogi, R. P., Richa, S. R. P., Singh, S. P. and H\"{a}der, D. P.: Photoprotective compounds from marine organisms, J Ind Microbiol Biotechnol, 37, 537-558, 2010.

\bibitem[Rastogi et al. (2014)]{LABEL}
Rastogi, R. P., Sinha, R. P., Moh, S. H., Lee, T. K., Kottuparambil, S., Kim, Y. J., Rhee, J. S., Choi, E. M., Brown, M. T., H\"{a}der, D. P. and Han, T.: Ultraviolet radiation and cyanobacteria, Journal of Photochemistry and Photobiology B: Biology, 141, 154-169, 2014.

\bibitem[\u{R}ezanka et al. (2004)]{LABEL}
\u{R}ezanka, T., Temina, M., Tolsikov, A. G. and Dembitsky, V. M.: Natural microbial UV radiation filters - mycosporine-like amino acids, Folia Microbiol, 49, 339-352, 2004. 

\bibitem[Richards et al. (2006)]{LABEL}
Richards, T. A., Dacks, J. B., Campbell, S. A., Blanchard, J. L., Foster, P. G., McLeod, R. and Roberts, C. W.: Evolutionary origins of the eukaryotic shikimate pathway: gene fusions, horizontal gene transfer, and endosymbiotic replacements, Eukaryot Cell, 5(9), 1517-1531, 2006.

\bibitem[Roeselers et al. (2007)]{LABEL}
Roeselers, G., Norris, T. B., Castenholz, R. W., Rysgaard, S., Glud, R. N., K\"{u}hl, M. and Muyzer, G.: Diversity of phototrophic bacteria in microbial mats from Arctic hot springs (Greenland), Environ. Microbiol., 9, 26-38, 2007.

\bibitem[Romera-Castillo et al. (2010)]{LABEL}
Romera-Castillo, C., Sarmento, H., \'{A}lvarez-Salgado, X. A., Gasol, J. M. and Marras\'{e}, C.: Production of chromophoric dissolved organic matter by marine phyto-plankton, Limnol. Oceanogr., 55(1), 446-454, 2010.

\bibitem[Rosic and Dove (2011)]{LABEL}
Rosic, N. N. and Dove, S.: Mycosporine-like amino acids from coral dinoflagellates, Appl Environ Microbiol., 77, 8478-8486, 2011.

\bibitem[Rossi and De Philippis (2015)]{LABEL}
Rossi, F. and De Philippis, R.: Role of Cyanobacterial Exopolysaccharides in Phototrophic Biofilms and in Complex Microbial Mats, Life (Basel), 5(2), 1218-1238, 2015.

\bibitem[Rossi et al. (2012)]{LABEL}
Rossi, F., Potrafka, R. M., Garcia-Pichel, F. and De Philippis, R.: The role of the exopolysaccharides in enhancing hydraulic conductivity of biological soil crusts, Soil Biol. Biochem., 46, 33-40, 2012.

\bibitem[R\"{0}ttgers and Koch (2012)]{LABEL}
R\"{o}ttgers, R. and Koch, B. P.: Spectroscopic detection of a ubiquitous dissolved pigment degradation product in subsurface waters of the global ocean, Biogeosciences, 9, 2585-2596, 2012.

\bibitem[Rowan (1989)]{LABEL}
Rowan, K. S.: Photosynthetic Pigments of Algae, Cambridge University Press, Cambridge, 112-210, 1989.

\bibitem[Rozema et al. (2002)]{LABEL}
Rozema, J., Bj\"{o}rn, L. O., Bornman, J. F., Gaberscik, A., H\"{a}der, D. P., Trost, T., Germ, M., Klisch, M., Gr\"{o}niger, A., Sinha, R. P., Lebert, M., He, Y. Y., Buffoni-Hall, R., de Bakker, N. V., van de Staaij, J. and Meijkamp, B. B.: The role of UV-B radiation in aquatic and terrestrial ecosystems - an experimental and functional analysis of the evolution of UV-absorbing compounds, J. Photochem. Photobiol. B., 66, 2-12, 2002.

\bibitem[Ruban et al. (2007)]{LABEL}
Ruban, A. V., Berera, R., Ilioaia, C., van Stokkum, I. H., Kennis, J. T., Pascal, A. A., van Amerongen, H., Robert, B., Horton, P. and van Grondelle, R.: Identification of a mechanism of photoprotective energy dissipation in higher plants, Nature, 450(7169), 575-578, 2007.

\bibitem[Sagan (1973)]{LABEL}
Sagan, C.: Ultraviolet selection pressure on the earliest organisms, J. Theor. Biol., 39, 195-200, 1973.

\bibitem[Sampedro (2011)]{LABEL}
Sampedro, D.: Computational exploration of natural sunscreens, Phys. Chem. Chem. Phys., 13, 5584-5586, 2011.

\bibitem[Schermann (2008)]{LABEL}
Schermann, J-P.: Spectroscopy and modeling of biomolecular building blocks, Amsterdam, Oxford: Elsevier, 2008.

\bibitem[Schmid et al. (2000)]{LABEL}
Schmid, D., Sch\"{u}rch, C. and Z\"{u}lli, F.: UV-A sunscreen from red algae for protection against premature skin aging, Cosmet. Toilet. Manuf. Worldw., 115, 139-143, 2000.

\bibitem[Schopf (1993)]{LABEL}
Schopf, J. W.: Microfossils of the Early Archean Apex chert: new evidence of the antiquity of life, Science, 260, 640-646, 1993.

\bibitem[Schopf and Packer (1987)]{LABEL}
Schopf, J. W. and Packer, B. M.: Early Archean (3.3-billion to 3.5-billion-year-old) microfossils from Warrawoona Group, Australia, Science, 237, 70-73, 1987.

\bibitem[Schopf et al. (2002)]{LABEL}
Schopf, J. W., Kudryavtsev, A. B., Agresti, D. G., Wdowiak, T. J. and Czaja, A. D.: Laser-Raman imagery of Earth's earliest fossils, Nature, 416, 73-76, 2002.

\bibitem[Shick and Dunlap (2002)]{LABEL}
Shick, J. M. and Dunlap, W. C.: Mycosporine-like amino acids and related gadusols: biosynthesis, accumulation, and UV protective functions in aquatic organisms, Annu Rev Plant Physiol., 64, 223-262, 2002.

\bibitem[Shick et al. (1999)]{LABEL}
Shick, J. M., Romaine-Lioud, S., Ferrier-Pag\`{e}s, C. and Gattuso, J. P.: Ultraviolet-B radiation stimulates shikimate pathway-dependent accumulation of mycosporine-like amino acids in the coral Stylophora pistillata despite decreases in its population of symbiotic dinoflagellates, Limnol Oceanogr, 44, 1667-1682, 1999.

\bibitem[Siegel et al. (2002)]{LABEL}
Siegel, D. A., Maritorena, S., Nelson, N. B., Hansell, D. A. and Lorenzi-Kayser, M.: Global distribution and dynamics of colored dissolved and detrital organic materials, J. Geophys. Res., 107(C12), 3228, 2002.

\bibitem[Singh et al. (2012)]{LABEL}
Singh, S. P., H\"{a}der, D. P. and Sinha, R. P.: Bioinformatics evidence for the transfer of mycosporine-like amino acid core (4-deoxygadusol) synthesizing gene from cyanobacteria to dinoflagellates and an attempt to mutate the same gene (YP\_324358) in Anabaena variabilis PCC 7937, Gene, 500(2), 155-63, 2012.

\bibitem[Singh et al. (2010)]{LABEL}
Singh, S. P., Klisch, M., Sinha, R. P. and H\"{a}der, D. P.: Genome mining of mycosporine-like amino acid (MAA) synthesizing and non-synthesizing cyanobacteria: A bioinformatics study, Genomics, 95(2), 120-128, 2010.

\bibitem[Sinha and H\"{a}der (2008)]{LABEL}
Sinha, R. P. and H\"{a}der, D. P.: UV-protectants in cyanobacteria, Plant Sci., 174, 278-289, 2008.

\bibitem[Sinha et al. (1999)]{LABEL}
Sinha, R. P., Klisch, M., Vaishampayan, A. and H\"{a}der, D. P.: Biochemical and spectroscopic characterization of the cyanobacterium Lyngbya sp. Inhabiting Mango (Mangifera indica) trees: presence of an ultraviolet-absorbing pigment, scytonemin, Acta Protozool, 38, 291-298, 1999.

\bibitem[Sinha et al. (2007)]{LABEL}
Sinha, R. P., Singh, S. P. and H\"{a}der, D. P.: Database on mycosporines and mycosporine-like amino acids (MAAs) in fungi, cyanobacteria, macroalgae, phytoplankton and animals, J. Photochem. Photobiol. B, 89, 29-35, 2007.

\bibitem[Sorrels et al. (2009)]{LABEL}
Sorrels, C. M., Proteau, P. J., and Gerwick, W. H.: Organization, evolution, and expression analysis of the biosynthetic gene cluster for scytonemin, a cyanobacterial UV-absorbing pigment, Appl. Environ. Microbiol., 75, 4861-4869, 2009.

\bibitem[Soule et al. (2009b)]{LABEL}
Soule, T., Garcia-Pichel, F., and Stout, V.: Gene expression patterns associated with the biosynthesis of the sunscreen scytoneminin Nostoc punctiforme ATCC29133 in response to UVA radiation, J. Bacteriol., 191, 4639-4646, 2009b.

\bibitem[Soule et al. (2009a)]{LABEL}
Soule, T., Palmer, K., Gao, Q., Potrafka, R. M., Stout, V. and Garcia-Pichel, F.: A comparative genomics approach to understanding the biosynthesis of the sunscreen scytonemin in cyanobacteria, BMC Genomics, 10, 336, 2009a.

\bibitem[Soule et al. (2007)]{LABEL}
Soule, T., Stout, V., Swingley, W. D., Meeks, J. C., and Garcia-Pichel, F.: Molecular genetics and genomic analysis of scytonemin biosynthesis in Nostoc punctiforme ATCC29133, J. Bacteriol., 189, 4465-4472, 2007.

\bibitem[Staleva et al. (2015)]{LABEL}
Staleva, H., Komenda, J., Shukla, M. K., \v{S}louf, V., Ka\v{n}a, R., Pol\'{i}vka, T. and Sobotka, R.: Mechanism of photoprotection in the cyanobacterial ancestor of plant antenna proteins, Nat Chem Biol., 11(4), 287-291, 2015.

\bibitem[Stanley (2008)]{LABEL}
Stanley, S. M.: Earth System History, 3rd edn., W. H. Freeman and Company, New York, NY, 263-287, 2008.

\bibitem[Starcevic et al. (2008)]{LABEL}
Starcevic, A., Akthar, S., Dunlap, W. C., Shick, J. M., Hranueli, D., Cullum, J. and Long, P. F.: Enzymes of the shikimic acid pathway encoded in the genome of a basal metazoan, Nematostella vectensis, have microbial origins, Proc. Natl. Acad. Sci. USA, 105, 2533-2537, 2008.

\bibitem[Steinberg et al. (2004)]{LABEL}
Steinberg, D. K., Nelson, N., Carlson, C. A. and Prusak, A. C.: Production of chromophoric dissolved organic matter (CDOM) in the open ocean by zooplankton and the colonial cyanobacterium Trichodesmium spp., Mar. Ecol. Prog. Ser., 267, 45-56, 2004.

\bibitem[Strome et al. (2015)]{LABEL}
Storme, J. Y., Golubic, S., Wilmotte, A., Kleinteich, J., Vel\'{a}zquez, D. and Javaux, E. J.: Raman Characterization of the UV-Protective Pigment Gloeocapsin and Its Role in the Survival of Cyanobacteria, Astrobiology, 15(10), 843-857, 2015.

\bibitem[Subramaniam et al. (1999)]{LABEL}
Subramaniam, A., Carpenter, E. J., Karentz, D. and Falkowski, P. G.: Bio-optical properties of the marine diazotrophic cyanobacteria Trichodesmium spp. I. Absorption and photosynthetic action spectra, Limnol Oceanogr, 44, 608-617, 1999.

\bibitem[Tice and Lowe (2004)]{LABEL}
Tice, M. M and Lowe, D. R.: Photosynthetic microbial mats in the 3,416-Myr-old ocean, Nature, 431(7008), 549-552, 2004. 

\bibitem[Tilstone et al. (2010)]{LABEL}
Tilstone, G., Airs, R., Martinez-Vicente, V., Widdicombe, C. and Llewellyn, C.: High concentrations of Mycosporine-like amino acids and coloured dissolved organic material in the sea surface microlayer off the Iberian Peninsula, Limnol. Oceanogr., 55(5), 1835-1850, 2010.

\bibitem[Trione and Leach (1969)]{LABEL}
Trione, E. J., and Leach, C. M.: Light-induced sporulation and sporogenic substances in fungi, Phytopathology, 59, 1077-1083, 1969.

\bibitem[Uemura et al. (1980)]{LABEL}
Uemura, D., Katayama, C., Wada, A. and Hirata, Y.: Crystal and molecular structure of palythene possessing a novel 360 nm chromophore, Chem. Lett., 6, 755-756, 1980. 

\bibitem[Ustin et al. (2009)]{LABEL}
Ustin, S. L., Valko, P. G., Kefauver, S. C., Santos, M. J., Zimpfer, J. F. and Smith, S. D.: Remote sensing of biological soil crust under simulated climate change manipulations in the Mojave Desert, Remote Sens. Environ., 113, 317-328, 2009.

\bibitem[Van Kranendonk et al. (2008)]{LABEL}
Van Kranendonk, M. J., Philippot, P., Lepot, K., Bodorkos, S. and Pirajno, F.: Geological setting of Earth's oldest fossils in the c. 3.5 Ga Dresser Formation, Pilbara craton, Western Australia, Precambr. Res., 167, 93-124, 2008.

\bibitem[Varnali and Edwards (2010)]{LABEL}
Varnali, T. and Edwards, H. G. M.: Ab initio calculations of scytonemin derivatives of relevance to extremophile characterization by Raman spectroscopy, Philosophical Transactions of the Royal Society A: Mathematical, Physical and Engineering Sciences, 368(1922), 3193-3203, 2010.

\bibitem[Varnali and Edwards (2014)]{LABEL}
Varnali, T. and Edwards, H. G. M.: Raman spectroscopic identification of scytonemin and its derivatives as key biomarkers in stressed environments, Phil. Trans. R. Soc. A., 372, 20140197, 2014.

\bibitem[Vernet and Whitehead (1996)]{LABEL}
Vernet, M. and Whitehead, K.: Release of ultraviolet-absorbing compounds by the red-tide dinoflagellate Lingulodinium polyedra, Mar Biol., 127, 35-44, 1996

\bibitem[Wada et al. (2015)]{LABEL}
Wada, N., Sakamoto, T. and Matsugo, S.: Mycosporine-Like Amino Acids and Their Derivatives as Natural Antioxidants, Antioxidants, 4, 603-646, 2015.

\bibitem[Waller et al. (2006)]{LABEL}
Waller, R. F., Slamovits, C. H. and Keeling, P. J.: Lateral gene transfer of a multigene region from cyanobacteria to dinoflagellates resulting in a novel plastid-targeted fusion protein, Mol. Biol. Evol., 23, 1437-1443, 2006.

\bibitem[Walter et al. (1980)]{LABEL}
Walter, M. R., Buick, R. and Dunlop, J. S. R.: Stromatolites 3,400-3,500 Myr old from the North Pole area, Western Australia, Nature, 284, 443-445, 1980.

\bibitem[Westall (2009)]{LABEL}
Westall, F.: Life on an anaerobic planet, Science, 323, 471-472, 2009.

\bibitem[Westall et al. (2006)]{LABEL}
Westall, F., de Ronde, C. E. J., Southam, G., Grassineau, N., Colas, M., Cockell, C. and Lammer, H.: Implications of a 3.472-3.333 Gyr-old subaerial microbial mat from the Barberton greenstone belt, South Africa for the UV environmental conditions on the early Earth, Philosophical Transactions of The Royal Society B, 361, 1857-1875, 2006.

\bibitem[Whitehead and Vernet (2000)]{LABEL}
Whitehead, K. and Vernet, M.: Influence of mycosporine-like amino acids (MAAs) on UV absorption by particulate and dissolved organic matter in La Jolla Bay, Limnol. Oceanogr., 45, 1788-1796, 2000. 

\bibitem[W\"{u}rfel, P. and Ruppel (1985)]{LABEL}
W\"{u}rfel, P. and Ruppel, W.: The flow equilibrium of a body in a radiation field, J. Phys. C: Solid State Phys. 18, 2987-3000, 1985.

\bibitem[Wynn-Williams et al. (1999)]{LABEL}
Wynn-Williams, D. D., Edwards, H. G. M. and Garcia-Pichel, F.: Functional biomolecules of Antarctic stromatolitic and endolithic cyanobacterial communities, Eur. J. Phycol., 34, 381-391, 1999.

\bibitem[Wynn-Williams et al. (2002)]{LABEL}
Wynn-Williams, D. D.,  Edwards, H. G. M.,  Newton, E. M., and Holder, J. M.: Pigmentation as a survival strategy for ancient and modern photosynthetic microbes under high ultraviolet stress on planetary surfaces, International Journal of Astrobiology, 1(1), 39-49, 2002.

\bibitem[Yamashita and Tanoue (2003)]{LABEL}
Yamashita, Y. and Tanoue, E.: Chemical characterization of protein-like fluorophores in DOM in relation to aromatic amino acids, Mar. Chem., 82, 255-271, 2003.

\end{thebibliography}
\end{document}